\documentclass[useAMS,usenatbib,fleqn]{mn2e}

\usepackage{graphicx}
\usepackage{amssymb}
\usepackage{subfigure}
\newcommand\msun{{\,M_\odot}}

\newcommand\zsun{{\rm \,Z_\odot}}
\newcommand\lsun{{\rm \,L_\odot}}
\newcommand{\unit}[1]{\ensuremath{\, \mathrm{#1}}}
\usepackage{aecompl}
\usepackage[T1]{fontenc}
\usepackage{threeparttable}
\usepackage{caption}
\usepackage{times}

\newcommand{\cMpc}{~\mbox{comoving}~\mbox{Mpc}}

\newcommand{\pc}{~\mbox{pc}}

\newcommand{\ckpc}{~\mbox{comoving}~\mbox{kpc}}

\newcommand{\cmci}{~\mbox{cm}^{-3}}

\newcommand{\Msunyri}{~\mbox{M}_{\odot}~\mbox{yr}^{-1}}

\title[Feedback-regulated first galaxy assembly]{The first galaxies: simulating their feedback-regulated assembly}
\author[Jeon et al.]{Myoungwon Jeon$^{1}$\thanks{E-mail:
myjeon@astro.as.utexas.edu},Volker Bromm$^{1}$, Andreas H. Pawlik$^{2}$ and Milo\v s Milosavljevi\'{c}$^{1}$\\
$^{1}$Department of Astronomy, University of Texas, Austin, TX 78712, USA \\ 
$^{2}$Max-Planck-Institut f\"ur Astrophysik, Karl-Schwarzschild-Strasse 1, 85748 Garching bei M\"unchen, Germany}

\begin{document}

\date{}

\pagerange{\pageref{firstpage}--\pageref{lastpage}} \pubyear{xxxx}

\maketitle
\topmargin-1cm

\label{firstpage}

\begin{abstract}

We investigate the formation of a galaxy reaching a virial mass of 
$\approx 10^8\msun$ at $z\approx10$ by carrying out a zoomed
radiation-hydrodynamical cosmological simulation. This simulation
traces Population~III (Pop~III) star formation, characterized by a
modestly top-heavy initial mass function (IMF), and considers stellar
feedback such as photoionization heating from Pop III and 
Population~II (Pop~II) stars, mechanical and chemical feedback from supernovae
(SNe), and X-ray feedback from accreting black holes (BHs) and
high-mass X-ray binaries (HMXBs). We self-consistently impose a
transition in star formation mode from top-heavy Pop III to low-mass
Pop~II, and find
that the star formation rate in the computational box is dominated by
Pop~III until $z\sim13$, and by Pop~II thereafter. 
The simulated galaxy experiences bursty
star formation, with a substantially reduced gas content due to
photoionization heating from Pop~III and Pop~II stars, together with
SN feedback. 
All the gas within the simulated galaxy is
metal-enriched above $10^{-5}\zsun$, such that there are no
remaining pockets of primordial gas. The
simulated galaxy has an estimated observed flux of 
$\sim 10^{-3} \unit{nJy}$,
which is too low to be detected by the James Webb Space Telescope
(JWST) without strong lensing amplification. We also show that our
simulated galaxy 
is similar in terms of stellar mass to Segue 2, the least
luminous dwarf known in the Local Group.
\end{abstract}


\section{Introduction}

An important goal of modern cosmology is to understand the formation and properties of early galaxies that formed a few 
hundred million years after the Big Bang \citep{Bromm2011, Loeb2013, Wiklind2013}. In the context of hierarchical 
cosmology, where small structures form first and evolve into bigger systems, dwarf galaxies with virial masses of 
$M_{\rm vir}=10^7-10^9\msun$ at $z\simeq 6-15$ were the basic building blocks of large galaxies seen today. 
Furthermore, it is thought that these galaxies played a pivotal role in reionizing the Universe, the global phase transition from 
an early neutral medium to a fully ionized one (for reviews see, e.g. \citealp{Barkana2007}; \citealp{Furlanetto2006}; 
\citealp{Robertson2010}). Specifically, the first galaxies were the drivers of the initial stages of reionization
(\citealp{Haardt2012}; \citealp{Shull2012}; \citealp{Robertson2013}). 
\par
Over the past decade, there have been 
enormous efforts to push the high redshift frontier to $z\simeq7$, detecting light from galaxies with a stellar mass of 
$M_{\ast}\gtrsim10^8\msun$ (e.g. \citealp{Bouwens2011}; \citealp{Oesch2012}; \citealp{Finkelstein2013};  
\citealp{Schenker2014}; \citealp{Stark2014}). To assess the observability of even smaller galaxies, and galaxies at 
even higher redshifts, we need to predict the properties of the first galaxies 
by pushing numerical simulations to new levels of physical realism and detail. Our goal here is to derive ab-initio 
predictions for key quantities, such as their stellar population mix, quantified by a possibly time-variable initial mass function (IMF), 
star formation rates (SFRs), metallicities, and resulting broad-band colour and recombination-line spectra (e.g. \citealp{Johnson2009}; 
\citealp{Pawlik2011a}; \citealp{Zackrisson2011}). A SFR above $0.1 \msun \unit{yr^{-1}}$, corresponding to $M_{\ast}\sim10^7\msun$, 
might be required for galaxies at $z\gtrsim 10$ to be observed with the {\it James Webb Space Telescope (JWST)} 
(e.g. \citealp{Inoue2011}; \citealp{Pawlik2013}). For {\it JWST} deep fields to reach even fainter systems, a magnification boost from 
gravitational lensing may be required \citep{Zackrisson2012, Zackrisson2014}. 
\par
The properties of such early galaxies were determined by the feedback from
preceding generations of stars (for a review see, 
e.g. \citealp{Ciardi2005}). The first generation of stars, the so-called Population~III (Pop~III), formed at $z\lesssim30$ 
in dark matter (DM) minihaloes with $M_{\rm vir}=10^5-10^6\msun$, predominately via ro-vibrationally excited molecular 
hydrogen ($\rm H_2$) cooling (e.g. \citealp{Haiman1996}; \citealp{Tegmark1997}; \citealp{Bromm2002}; \citealp{Yoshida2003}). 
Radiation from Pop~III stars dramatically altered the gas within their host minihaloes, through photoionization, photoheating, 
and photoevaporation (e.g. \citealp{Kitayama2004}; \citealp{Whalen2004}; \citealp{Alvarez2006}; \citealp{Yoshida2007a}; 
\citealp{Wise2008}). Once a Pop~III star died as a supernova (SN), heavy elements were dispersed, enriching the intra-host 
and intergalactic medium (IGM), thus initiating the prolonged process of chemical evolution (reviewed in \citealp{Karlsson2013}).
The resulting SN feedback from the first stars greatly affected the gas in the shallow potential wells of the host systems
\citep[e.g.][]{Greif2007, Greif2010, Whalen2008}. The more distant, diffuse IGM was heated as well by X-rays emitted by accreting 
single black holes (BHs), or high-mass X-ray binaries (HMXBs), both remnants of Pop~III stars (e.g. \citealp{Glover2003}; 
\citealp{Kuhlen2005}; \citealp{Milos2009a,Milos2009b}; \citealp{Alvarez2009}; \citealp{Mirabel2011};
\citealp{Wheeler2011}; \citealp{Hummel2014}; \citealp{Jeon2012,Jeon2014a}; \citealp{Xu2014}).
\par
The critical role of feedback from stars in massive galaxies with $M_{\rm vir}\sim10^9-10^{13}\msun$ is also emphasized by many 
authors (e.g. \citealp{Schaye2010}; \citealp{Stinson2013}; \citealp{Kim2013}; \citealp{Wise2014}). Photoionization, radiation pressure, 
and SN feedback operate in a non-linear, coupled fashion, such that considering a full treatment of the radiation-hydrodynamics and 
time dependence of these feedback mechanisms is necessary to reproduce the observed histories of star formation, including their stochastic nature, particularly in dwarf 
galaxies (e.g. \citealp{Pawlik2009}; \citealp{Hopkins2013}). However, due to the prohibitive computational cost related to the baryonic physics, simulating the assembly of massive 
galaxies from first principles is not feasible. The first dwarf galaxies with virial masses $M_{\rm vir}\lesssim10^9$, where such ab-initio 
modeling is coming within reach, thus provide us with ideal laboratories to develop a detailed theory of feedback-regulated galaxy formation.
\par
The birth of the first galaxies is often defined by theorists as the emergence of atomic cooling haloes, systems assembled at $z\lesssim15$ 
with masses of $M_{\rm vir}\gtrsim5\times10^{7}\msun$, within which the primordial gas was able to cool via atomic hydrogen lines instead 
of molecular hydrogen \citep[e.g.][]{Oh2002, Bromm2003a}. The deeper potential wells of these haloes allowed them to sustain self-regulated 
star formation in a multi-phase medium by holding on to the gas driven out by stellar feedback. Another feature that renders the atomic cooling 
haloes viable first galaxy candidates is the presence of supersonic turbulence, which is believed to play an important role in present-day star 
formation \citep{McKee2007}. Such turbulence is expected to develop during the build-up of the first galaxies due to the inflow of cold gas along 
cosmic filaments (e.g. \citealp{Wise2007}; \citealp{Greif2008}; \citealp{SSC2012}).  
\par
The relative importance of Pop~III stellar feedback on first galaxy formation is determined by the mass range of Pop~III stars 
(e.g. \citealp{Bromm2013}). Stars with initial masses in the range of $40\msun-140\msun$, or larger than $260\msun$, are expected to directly 
collapse into BHs, whereas they undergo pair-instability supernovae (PISNe) in the range of $140\msun-260\msun$ (\citealp{Heger2002}). 
The PISN range can be extended down to $\sim85\msun$ if the progenitor is rapidly rotating (\citealp{Chatzopoulos2012,Yoon2012}). 
Finally, stars with masses between $8\msun$ and $40\msun$ likely died as conventional core-collapse SNe (CCSNe), or energetic 
hypernovae in case of high rotation rates (e.g. \citealp{Umeda2005}). 
\par
Earlier studies had predicted that the first stars formed in isolation,   
with a characteristic mass of $\sim100\msun$ (e.g. \citealp{Abel2002}; \citealp{Bromm2002}; \citealp{Yoshida2006}). 
However, recent high-resolution simulations, exploring the high densities encountered in protostellar disks, have suggested that the 
primordial cloud can further fragment, giving rise to the possibility of binary or multiple systems \citep{Turk2009,Stacy2010,Clark2011a,
Prieto2011,Smith2011,Greif2011,Greif2012,Stacy2012,Dopcke2013}. Radiation-hydrodynamical simulations that take into account the 
feedback from the protostellar radiation suggest that the first stars typically reached masses of a few $\sim 10\msun$
\citep[e.g.][]{Hosokawa2011,Stacy2012,Hirano2014,Susa2014}. If this is indeed the case, the feedback from primordial stars would have 
been less disruptive than previously thought (\citealp{Kitayama2005}; \citealp{Whalen2008}; \citealp{Chiaki2013}). Specifically, PISN events 
would have been less frequent, with the majority of Pop~III stars now expected to die as CCSNe. It is therefore important to revisit the assembly 
process of the first galaxies, taking this change in characteristic Pop~III mass scale into account.
\par
Most previous studies have been focusing on the feedback from massive Pop~III stars ($\gtrsim100\msun$) on first 
galaxy formation \citep[e.g.][]{Wise2008,Greif2010,Wise2012,Pawlik2013,Muratov2013,Xu2013,Wise2014}. One prominent result is that 
even a single PISN with an explosion energy of $E_{\rm SN}\sim10^{52}$~erg, and a metal yield of $y\sim0.5$, has a dramatic impact on 
subsequent evolution, evacuating the gas within the host halo, and preventing further star formation for about a Hubble time, $\sim$300\,Myr 
at $z\sim10$. As the ejected metals fall back into the emerging galaxy, the central gas is enriched up to $Z\gtrsim10^{-3.5}\zsun$ 
(e.g. \citealp{Wise2008}; \citealp{Greif2010}). On the other hand, the recovery from the feedback of less massive Pop~III stars ($\lesssim40\msun$), 
followed by CCSNe with $E_{\rm SN}\sim10^{51}$ and $y=0.05$, was rather prompt, with recovery timescales of only a few $\sim10$\,Myr \citep{Jeon2014b}. 
Such prompt recovery allowed the first galaxies to experience multiple episodes of star formation prior to their assembly (\citealp{Ritter2012,Ritter2014}). 
Related to this weakened feedback, the star-forming gas inside the first galaxies was possibly enriched by multiple SNe. The resulting metal enrichment 
was sufficient to affect a transition of the star formation mode from Pop~III to normal, Population~II (Pop~II) stars. Thus, regions that hosted moderate 
mass Pop~III stars would locally undergo the transition to (low-mass dominated) Pop~II star formation earlier on.
\par 
The minimum (``critical'') metallicity for the Pop~III/Pop~II transition depends on which cooling is responsible for the gas fragmentation 
(e.g. \citealp{Omukai2000}; \citealp{Bromm2001b}; \citealp{Schneider2002}; \citealp{BrommLoeb2003}; \citealp{Omukai2005}; 
\citealp{Schneider2010}; \citealp{Schneider2012};  \citealp{Chiaki2014}; \citealp{Ji2014}). Fine-structure line cooling, mainly via C~II and 
O~I, above $Z\sim10^{-3.5}\zsun$, leads to vigorous fragmentation of the gas cloud, giving rise to the formation of stellar clusters. Further fragmentation 
can be achieved at very high densities, $n_{\rm H}\gtrsim10^{16}\unit{cm^{-3}}$, via dust continuum cooling if the initial gas-to-dust ratio exceeds 
$\it{D}_{\rm crit}=\rm [2.6-6.3]\times10^{-9}$ (\citealp{Schneider2012}; \citealp{Chiaki2014}). Globally, the Pop~III/Pop~II transition is a gradual 
process, as the ejected metals are brought back into virialized host haloes, keeping a large fraction of the IGM still pristine. Therefore, those two 
populations co-existed, such that Pop~III star formation was still possible even at $z\sim 6$ (e.g. \citealp{Trenti2009, Maio2010, Maio2011, 
Scannapieco2003, Wise2012, Johnson2013}).
\par 
Here, we present the results of a cosmological zoom-in, high-resolution radiation-hydrodynamics simulation, following the first galaxy 
assembly process from first principles. We consider realistic descriptions of Pop~III/Pop~II star formation, photoionization and photoheating 
from stars, the mechanical and chemical feedback from PISNe and CCSNe, and the X-ray feedback from accreting BHs and HMXBs. Guided by the 
results from recent investigations of primordial star formation, we in particular explore the implications of a somewhat less top-heavy Pop~III IMF, 
with a characteristic mass of a few $\sim10\msun$, giving rise to a large fraction of CCSNe rather than PISNe.
\par  
The outline of the paper is as follows. Our numerical methodology is described in Section~2, and the simulation results are presented 
in Section~3. Finally, our main findings are summarized in Section~4.
For consistency, all distances are expressed in physical (proper) units unless noted otherwise.
\label{Sec:Intro}

\section{Numerical methodology}
\label{Sec:Metho}
\subsection{Gravity, hydrodynamics, and chemistry}
We have performed our simulations using a modified version of the $N$-body/TreePM Smoothed Particle
Hydrodynamics (SPH) code GADGET (\citealp{Springel2001}; \citealp{Springel2005}). 
The initial conditions within a cubic volume $1 \cMpc$ on a side are generated by assuming a 
$\Lambda$CDM cosmology with a matter density parameter of $\Omega_{\rm m}=1-\Omega_{\Lambda}=0.3$,
baryon density $\Omega_{\rm b}=0.04$, present-day Hubble expansion rate $H_0
= 70\unit{km\, s^{-1} Mpc^{-1}}$, spectral index $n_{\rm s}=1.0$, and
normalization $\sigma_8=0.9$. The initial conditions are hierarchically refined by a consecutive zoom-in technique, 
rendering the masses of dark matter (DM) and SPH particles in the highest resolution region with an approximate 
linear size of $300 \ckpc$ $m_{\rm DM} \approx 33\msun$ and $m_{\rm SPH} \approx 5 \msun$, respectively. 
We adopt the gravitational softening length $\epsilon_{\rm soft}=70$ comoving pc for both DM and baryonic particles.
We start our simulations at $z \approx 100$ and follow the cosmic evolution of the DM and gas until the assembly process 
of the first galaxy has been completed at $z\approx10$.
\par
We consider all relevant primordial chemistry and cooling processes. Specifically, we include H and He 
collisional ionization, excitation and recombination cooling, bremsstrahlung, inverse Compton cooling, and collisional
excitation cooling of $\rm H_2$ and HD. The code self-consistently solves the non-equilibrium chemistry
for 9 abundances ($\rm H, H^{+}, H^{-}, H_{2}, H^{+}_2 , He, He^{+}, He^{++},$ and $\rm e^{-}$), as well as for the three
deuterium species $\rm D, D^{+}$, and HD.  Additionally, metal-enriched gas can also be cooled by metal species, specifically 
C, O, and Si, with solar relative abundances. For the metal cooling rates, we adopt the results of  \citet{Glover2007}. 
The chemical network comprises the key species, $\rm C, C^{+}, O, O^{+}, Si, Si^{+}$ and $\rm Si^{++}$. We 
assume that all the metals are in gas phase atoms or ions, and thus we do not consider 
dust cooling as the gas densities explored in this work are far below the threshold value above which it  
starts to become important.

\subsection{Star formation physics}
\subsubsection{Population~III}
For the star formation, we do not form stars by following the evolution from the initial protostars to the final mass 
(e.g. \citealp{Hirano2014}), rather we adopt a sink algorithm (\citealp{Johnson2006}), where stars are represented by 
sink particles. Once a gas particle exceeds the threshold density, $n_{\rm H, max}=10^4 \cmci$, the highest-density SPH 
particle is converted into a collisionless sink particle, subsequently accreting neighboring gas particles until the mass of 
the Pop~III star is reached. The masses of Pop~III stars are randomly sampled from an IMF with a functional form of 
\begin{equation}
\frac{dN}{d \ln M} \propto M^{-1.3} \exp{\left[ -\left(\frac{M_{\rm char}}{M}\right)^{1.6}\right]},
\end{equation}
where $M_{\rm char}=20\msun$ is the characteristic mass. Above $M_{\rm char}$, it behaves as a Salpeter-like IMF, but 
is exponentially cutoff below that mass (e.g. \citealp{Chabrier2003}; \citealp{Wise2012}). 
Once a sink particle forms, we prevent the subsequent star formation in a sphere with radius 30 pc centered on the sink particle for $\sim1$ Myr. This is because it 
takes time for the hydrodynamical shocks with a velocity $\sim$30 km/s, generated by photoheating to sweep up and 
evacuate the central gas, reducing the central gas densities. Without such temporary suppression in star formation, 
multiple stars might form at the same location at the same time. This would correspond to fragmentation within the SPH smoothing kernel, which is not resolved, and would thus be physically unreliably. We allow sink particles to merge once their distance 
falls below 1\,pc, similar to the baryonic resolution scale $l_{\rm res} \equiv [( 3X M_{\rm res})/(4 \pi n_{\rm H, max}
  m_{\rm H})]^{1/3} \approx 0.5 \pc$, where $X=0.76$ is the hydrogen mass fraction. Here, the baryonic mass resolution is 
  $M_{\rm res} \equiv N_{\rm ngb} m_{\rm SPH}\approx 240\msun$, where $N_{\rm ngb} = 48$ is the number of particles in 
  the SPH smoothing kernel \citep{Bate1997}. The new positions and velocities of the merged sinks are 
  computed by mass-weighted averaging.  
\par
\subsubsection{Population~II}
Pop~II stars are formed out of the gas cloud which is metal-enriched by the previous generation of stars. The star formation 
recipe is similar to that of Pop~III star formation except for one more condition: if the metallicity of a gas 
particle, eligible for star formation, exceeds the characteristic metallicity, $Z_{\rm crit}=10^{-3.5}\zsun$, for the transition from 
Pop~III to Pop~II, we assume that a newly formed sink particle represents a Pop~II star cluster. Therefore, once a SPH particle 
satisfies those two criteria, $n_{\rm H, max}=10^4 \cmci$ and $Z_{\rm crit}=10^{-3.5}\zsun$, we immediately create an effective sink particle with a mass of $M_{\rm Pop~II}=500\msun$, by accreting 
surrounding gas particles. The mass of a Pop~II cluster $M_{\rm Pop~II}=500\msun$ is 
chosen to host one massive star $\sim20\msun$, given a Salpeter IMF, $dN/d\log m\approx m^{-\alpha}$, with a slope 
$\alpha=1.35$, which is from \citet{Schaerer2003} where they provide photoionization rates derived from an assumed IMF 
for a Pop~II cluster. In the presence of a $20\msun$ star, further star formation inside 
of the cluster after the initial burst is truncated by the photoheating of the star (\citealp{SSC2014b}). 
\par

\subsection{Feedback physics}
\subsubsection{Photoionization feedback}
Once a star forms, the star particle emits ionizing radiation over its lifetime with a black-body spectrum. 
We use the radiative transfer (RT) code TRAPHIC to transport ionizing photons \citep{Pawlik2008,Pawlik2011}. 
In simulations with TRAPHIC, photon packets from radiation sources are transferred along the irregular, 
spatially adaptive grid set by SPH particles in a photon-conserving manner through the simulation box. 
Thanks to a photon packet merging technique in TRAPHIC, the computational cost is independent of the 
number of ionizing radiation sources, which enables large simulations considering many sources, and as presented here. 
The RT is controlled by a set of parameters, which we choose identical 
to those employed in \citet{Jeon2014a}, except that we here use two frequency bins, $N_{\nu}=2$ instead of 
$N_{\nu}=4$, with bounding energies located at [13.6 eV, 400 eV, 10 keV]. While the main focus of 
\citet{Jeon2014a} is to investigate the impact of X-rays on the gas in haloes and IGM, requiring to track high 
energy photons denoted by the high characteristic frequencies, in this work we focus on the 
stellar feedback impact. To do so, two frequency bins, $N_{\nu}=2$, are sufficient to represent the stellar 
spectrum. We refer the reader to the original papers \citep{Pawlik2008,Pawlik2011} for further details 
concerning RT methods and to \citet{Jeon2014a} for the RT parameters we use in this work. The RT 
computation is coupled to the hydrodynamical evolution by passing the photoionization, photoheating, 
and photodissociation rates computed by the RT to the non-equilibrium solver for the chemical and thermal 
evolution of the gas.
\par
For the ionizing photon rates  of Pop~III stars and their lifetimes, we employ polynomial fits (\citealp{Schaerer2003}) as a function 
of initial mass of a star, $\log_{10} \dot{N}_{\rm ion}=43.61+4.90 x-0.83 x^2$ and 
$\log_{10} t_{\ast}=9.785-3.759x+1.413x^2-0.186x^3$, respectively, where $x=\log_{10}(M_{\rm Pop~III}/\msun)$. 
For Pop~II clusters, we use the value from \citet{Schaerer2003} that provides the properties of integrated stellar 
populations at various metallicities. In particular, we fixed the ionizing photon rate for the initial $\sim3$ Myr at 
$\dot{N}_{\rm ion}=10^{47} \rm s^{-1} \msun^{-1}$, which is derived by averaging over a Salpeter IMF with a 
slope of $\alpha=1.35$ and a metallicity of $Z=5\times10^{-3}\zsun$ in the mass range of [1$\msun$, 150$\msun$]. 
For the following evolution of the rates, we use a polynomial fit (\citealp{Schaerer2003}), $\log_{10} \dot{N}_{\rm ion}=48.13-0.42t+0.01t^2 \unit{s^{-1}\msun^{-1}}$, 
where $t$ is the elapsed time in Myr after Pop~II cluster formation. The emission from Pop~II stars lasts for 20 Myr, 
corresponding to the typical lifetime of OB stars in the cluster.
\par
In addition to the ionizing radiation from individual Pop~III stars and Pop~II clusters, we also track the Lyman-Werner
(LW) radiation from both populations that dissociates molecular hydrogen ($\rm H_2$) and deuterated hydrogen (HD). 
This is treated in the optical thin limit with a self-shielding correction (\citealp{Wolcott2011b}). The rate at which LW photons 
are emitted by Pop~III stars is fit using $\log_{10} \dot{N}_{\rm LW}=44.03+4.59x-0.77x^2$ (\citealp{Schaerer2002}). 
We only allow stars to form within the highly resolved region corresponding to a linear size of $
\approx$ 300 comoving kpc and do not follow the propagation of photons outside the region. For simplicity, we do not consider a
global UV background here.
\par

\begin{figure}
  \includegraphics[width=90mm]{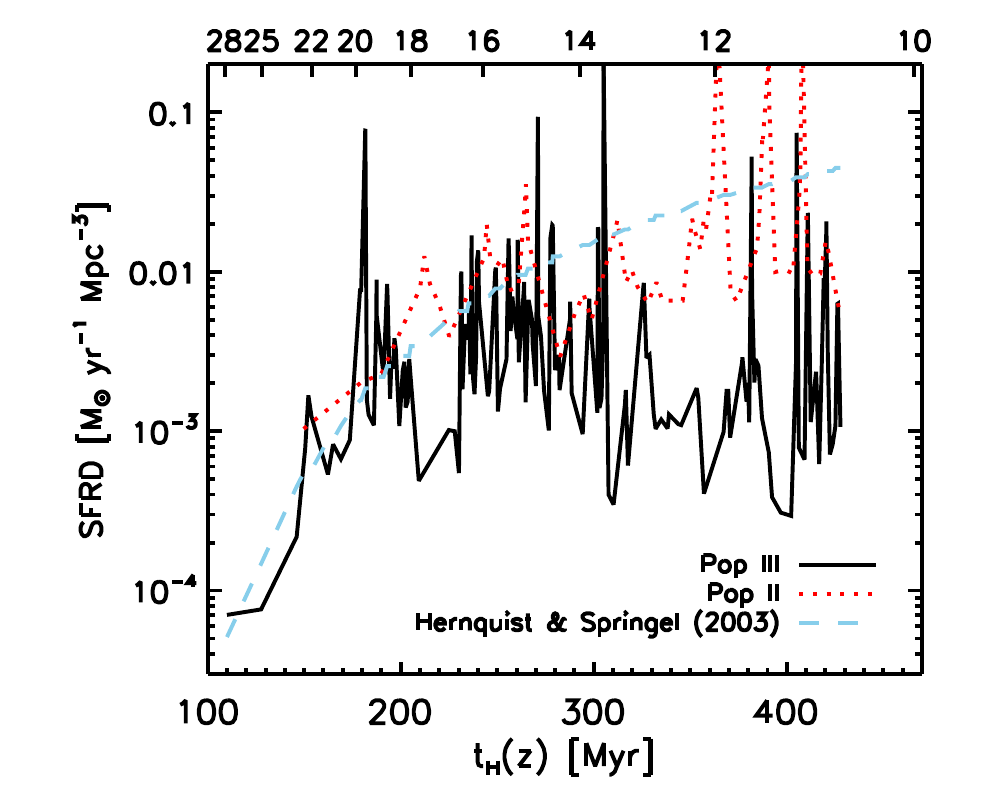}
     \caption{The history of comoving star formation rate density (SFRD) in the simulated region for 
     Pop~III (black solid line) and Pop~II (red dotted line) stars. For comparison, we also show the analytic 
     fitting formula, derived by Hernquist \& Springel (2003), for higher-mass systems where 
     atomic cooling is active (light blue-dashed line). Earlier on, at $z>13$, Pop~III 
     star formation dominates, but as the gas is metal-enriched, the Pop~II
     mode becomes dominant below $z\approx13$.
      \label{Fig:SFRD}}
\end{figure}

\subsubsection{Supernova feedback}
The SN explosion energy is inserted as thermal energy into the neighboring SPH particles, 
$\bar{N}_{\rm neigh}=32$, a subset of the particles within the SPH smoothing kernel, $N_{\rm neigh}$, 
around a sink particle. To ensure high accuracy and energy conservation, we use a timestep-limiter, with which the ratio of timesteps of neighboring SPH particles 
cannot be larger than 4 (\citealp{Saitoh2009}, \citealp{Durier2012}). Additionally, using a timestep-updater, implemented 
by \citet{Durier2012}, all neighboring particles, $\bar{N}_{\rm neigh}$, around an exploding sink particle become 
active particles at the time of SN energy injection, enabling an immediate hydrodynamical response.    
\par
At the end of a Pop~III star's lifetime, the fate of the Pop~III star is determined by its initial mass. 
For instance, Pop~III stars with masses between $9\msun$ and $40\msun$ are expected to end their 
lives as conventional core collapse SNe. Here, we use the fixed CCSN energy of $E_{\rm SN}=10^{51}$ 
ergs and metal yield, $y=0.05$, for Pop~III progenitors. We should mention that the CCSN explosion energies 
and the metal yields can vary depending on the progenitor's mass (e.g. \citealp{Heger2010}). In particular, 
under the existence of rotation of Pop~III progenitor in a mass range of $25-40\msun$, a much stronger explosion 
may be expected, likely resulting in a hypernova (e.g. \citealp{Fryer2000}; \citealp{Nomoto2006}). However, due to 
the uncertainty in the degree of spin of Pop~III stars, which is subject to stellar winds, gravitational and 
hydromagnetic instabilities (e.g. \citealp{Stacy2011}), we only consider conventional CCSNe by fixing the explosion 
energy and metal yield.  
\par
For the highly energetic PISNe from Pop~III stars with masses in 
the range of $140\msun \lesssim M_{\ast} \lesssim 260\msun$ (\citealp{Heger2002}), we adopt the SN 
energy of $10^{52}$ ergs and $y=0.5$. Out of a Pop~II cluster with a mass of $500\msun$, a total of 9 stars more 
massive than $8\msun$, above which the stars are eligible to explode as CCSNe, are expected to form. We 
simultaneously trigger 9 SNe explosions at the end of the lifetime of a Pop~II cluster, injecting an energy of 
$9\times10^{51}$ ergs with an IMF-averaged effective metal yield $y_{\rm eff}=0.005$ onto the neighboring 
particles around the Pop~II sink particle.
\par
Initially, the metals from SN explosions are evenly distributed among the neighboring SPH particles, implying 
initial metallicities, 
\begin{equation}
Z_i = \frac{m_{\rm metal,i}}{m_{\rm SPH}+m_{\rm metal,i}},
\end{equation}
where $m_{\rm metal,i}=M_{\ast}y/\bar{N}_{\rm neigh}$ for Pop~III stars and 
$m_{\rm metal,i}=M_{\rm Pop~II}y_{\rm eff}/\bar{N}_{\rm neigh}$ for Pop~II clusters. 
The lack of mass flux between SPH particles means that metals are transported ballistically in an 
SPH scheme (e.g. \citealp{Wiersma2009}). Physically, the mixing of metals is achieved through an unresolved 
turbulent cascade from large scales to small scales, which is unresolved. Therefore, often, a metal 
mixing is implemented by subgrid models in which physical processes below the smallest resolved scale are 
determined by physical quantities at resolved scales (e.g. \citealp{Schmidt2011}). We adopt a 
diffusion-based method implemented by \citet{Greif2009} which assume that the mixing 
efficiency on unresolved scales is determined by the physical properties on the scale of the SPH 
smoothing kernel (\citealp{Klessen2003}). Consequently, the metal diffusion rate is set by the local diffusion coefficient, 
$D=2$ $\rho$ $\tilde{v}$ $\tilde{l}$, where the length scale, $\tilde{l}$, is comparable to the smoothing length of 
the SPH kernel, $\tilde{l}=h$, and $\rho$ is the gas density. The velocity dispersion within the kernel, 
$\tilde{v}$, is given by
\begin{equation}
\tilde{v}_i^2 = \frac{1}{N_{\rm ngb}} \sum_j |v_i-v_j|^2.
\end{equation}
Here, $v_i$ and $v_j$ are the velocities of particles $i$ and $j$ within the kernel.
\par
\subsubsection{Black hole feedback}
Pop~III stars with masses between $40\msun$ and $140\msun$, or larger than $260\msun$ will directly 
collapse into BHs (\citealp{Heger2003}). The BHs then grow by Bondi-Hoyle accretion (\citealp{Bondi1944}), assuming 
that a fraction $\epsilon=0.1$ of the accreted mass is converted into ionizing radiation. 
Such stellar mass BHs accreting diffuse gas are called miniquasars (\citealp{Kuhlen2005}). The ionizing 
luminosities of miniquasars are determined by normalizing the total accretion luminosity 
 \begin{eqnarray}
L_{\rm BH} &\equiv &  \int_{0}^{10 {\rm keV}/h_{\rm P}} {L_{\nu} d\nu} = \frac{\epsilon}{1-\epsilon}\dot{M}_{\rm BH}c^2,
\end{eqnarray}
where $h_{\rm P}$ is Planck's constant and c is the speed of light. The BH (sink) particles grow in mass by accreting 
surrounding gas according to the accretion rate, $\dot{M}_{\rm BH} = (1-\epsilon)\dot{M}_{\rm acc}$, which is computed as  
\begin{equation}
\dot{M}_{\rm acc} = \frac{4 \pi G^2 M_{\rm BH}^2 \rho}{(c^2_{\rm s}+v^2_{\rm vel})^{3/2}}.
\end{equation}
Here, $c_{\rm s}$ is the sound speed, $\rho$ the gas density, $M_{\rm BH}$ the BH mass, and $v_{\rm rel}$ the relative 
velocity of the BH to the surrounding gas. Note that Bondi-Hoyle accretion is thus a numerical input, or recipe, and not an independent simulation result. The estimated rate is an upper limit. In reality, 
the true accretion rates are likely smaller than the nominal Bondi-Hoyle value, if feedback from radiation pressure is considered
(e.g. \citealp{Milos2009a,Milos2009b}; \citealp{Park2012}). In addition, at later time as the gas in an 
atomic cooling halo becomes turbulent, the vorticity of the turbulent gas can reduce the accretion onto the BH (\citealp{Krumholz2005}). Motivated by observations at low redshifts, we take the same form of a spectral 
energy distribution of the emerging radiation from an accreting single black hole, which is characterized by a thermal 
multi-color disk (e.g. \citealp{Pringle1981}; \citealp{Mitsuda1984}) at frequencies lower than 0.2 keV$/h_{\rm P}$ and 
a non-thermal power law component at higher frequencies (e.g. \citealp{Kuhlen2005}). For the details, we refer the 
reader to \citet{Jeon2014a}.
\par
Further, we assume that half of the black holes are 
formed in a binary system and then every third of those evolve into a HMXB after the primary 
turned into a BH (\citealp{Power2009}). For the HMXB phase of duration $\Delta t_{\rm HMXB}=2$ Myr, corresponding to the 
typical main-sequence lifetime of the donor stellar companion (e.g. \citealp{Belczynski2012}), we assume an HMXB luminosity 
equal to the Eddington luminosity, 
\begin{eqnarray}
L_{\rm HMXB} &\equiv &  \int_{0}^{10 {\rm keV}/h_{\rm P}} {L_{\nu} d\nu} = L_{\rm Edd} \\
&=& 1.4 \times 10^{40} {\rm erg \hspace{0.1cm} s^{-1}}
\left( \frac{M_{\rm BH}}{100 \msun} \right),
\end{eqnarray}
corresponding to accretion of gas from the companion at a rate
$2.2 \times 10^{-6} \Msunyri (M_{\rm BH} / 100 M_{\odot})$. The X-ray radiation is followed using RT. 
We also take into account the secondary 
ionizations, which are mainly done by high energy 
electrons left behind the first ionizations of hydrogens by hard energy photons emitted 
from an accreting BH or a HMXB. We use the energy dependent fits to the secondary 
ionization and heating fractions provided by \citet{Ricotti2002}.

\begin{figure*}
  \includegraphics[width=120mm]{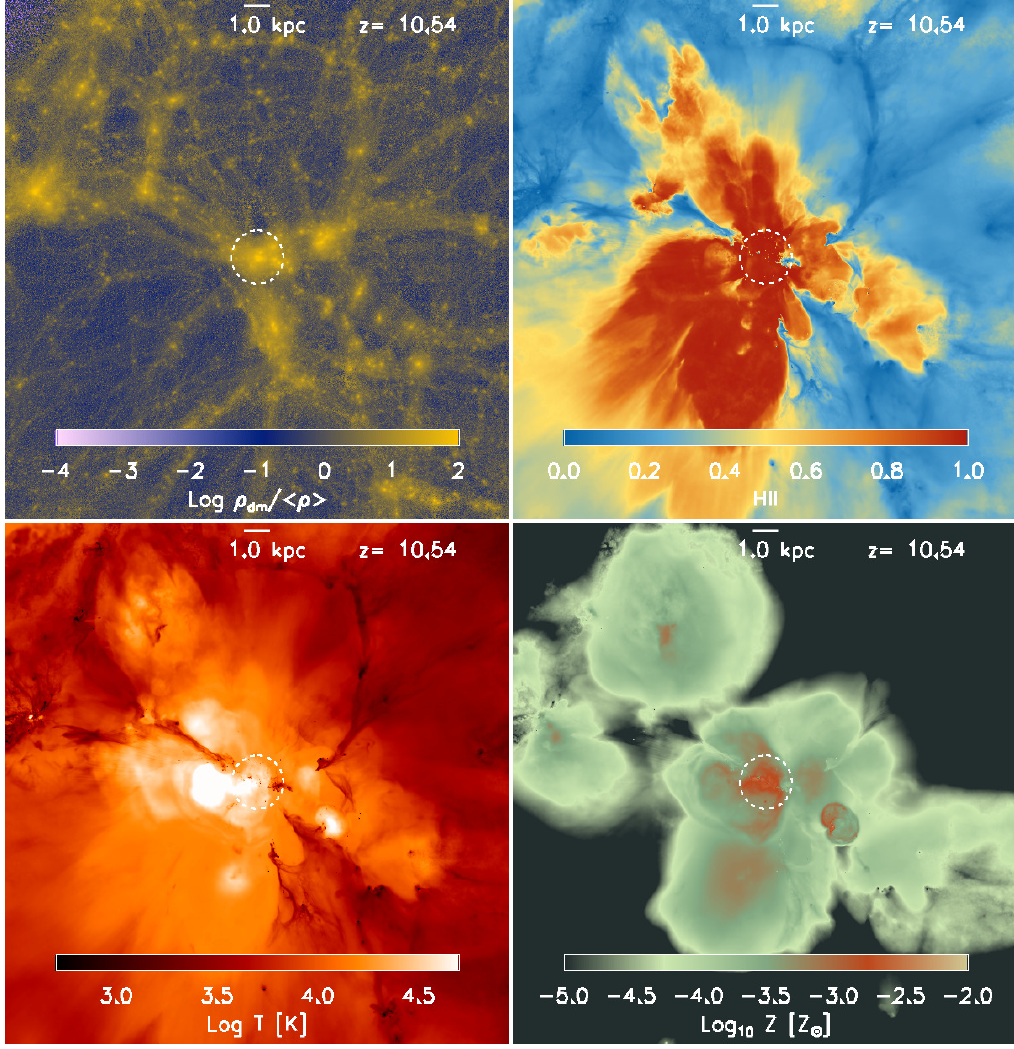}
     \caption{Morphology in the highly resolved region at $z\approx10.5$. {\it Clockwise from upper left}: Dark matter overdensity, 
     H~II fraction, gas metallicity, and gas temperature, averaged along the line of sight within 
     the central $\simeq350$~kpc (comoving). The central white circles denote the size of the 
     virial radius, $r_{\rm vir}\approx1.1$ kpc, of the emerging target galaxy. The target halo grows in mass through 
     mergers with neighboring haloes and accretion along the filaments of the cosmic web. The gas in haloes that are undergoing star formation 
     is evacuated and heated, and metal-enriched by stellar feedback. 
     The main metal contributors to the diffuse IGM are two energetic PISNe, formed at 
     $z\approx16$ and $z\approx15.5$, and ejecting a total of $\sim160\msun$ in heavy elements into the IGM. \label{Fig:snap_large}}
\end{figure*}

\begin{figure*}
  \includegraphics[width=160mm]{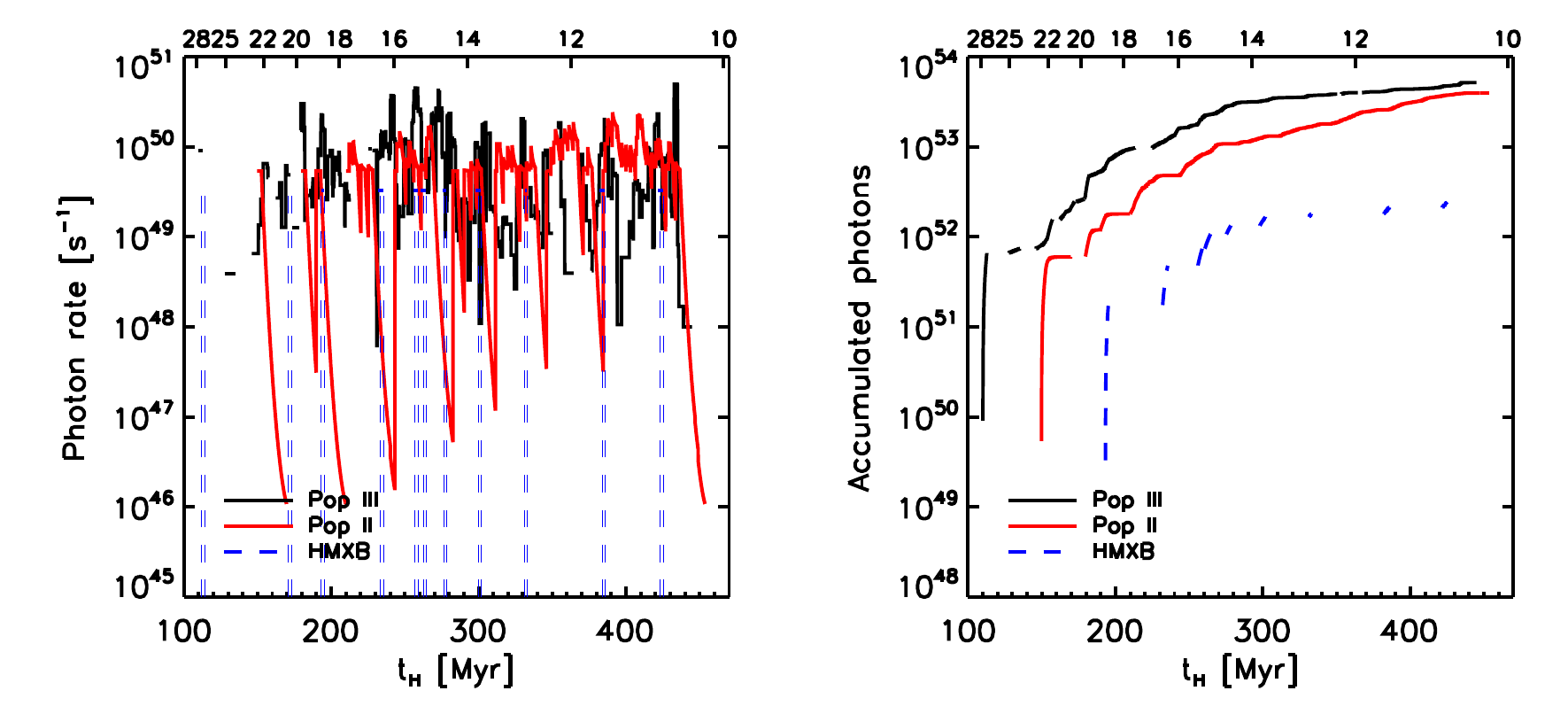}
     \caption{The evolution of ionizing photon rates ({\it left}) and total accumulated photons ({\it right}) 
     from the Pop~III stars, Pop~II clusters, and HMXBs. The contribution 
     from Pop~III stars dominates the total ionizing photon budget up to $z\approx10.5$, but Pop~II clusters are about to overtake it, as
     the global star formation mode changes to Pop~II star formation.
     The total photon production from the 11 HMXBs formed during the run is negligible, mainly due to our assumption here of
     a very brief interval of 2~Myr, over which the X-ray source is active. This interval, however, is uncertain, and could be larger. If so, the relative HMXB contribution to the ionizing photon budget could be more significant. \label{Fig:photons}}
\end{figure*}

\section{Simulation results}
In the following, we present our results. First, in Section 3.1, we discuss the global properties of the IGM 
and of haloes within the highly resolved region and their evolution to $z\approx10$. In Section 3.2, we 
focus on the assembly process of the central halo hosting the emerging galaxy. Then, based on the 
physical quantities derived from this galaxy, we compare the properties of the first galaxy 
with those of the local counterparts in Section 3.3. Finally, in Section 3.4, we discuss the detectability of the simulated first galaxy.

\subsection{Global evolution}
\subsubsection{Star formation history}
In the simulation, the first Pop~III star, with a randomly assigned mass of $100\msun$, forms at 
$z\approx28$ out of the primordial gas in a minihalo with mass $M_{\rm vir}\sim5\times10^5\msun$. After its
short lifetime of 2.7 Myr, the star ends up as a BH in a binary system, forming a HMXB. 
Over a brief interval of $\Delta t_{\rm HMXB}=2$ Myr, X-rays emitted 
from the HMXB further heat the surrounding gas, which was already heated to 
$\sim10^4$ K by the Pop~III progenitor, and they heat the distant IGM to a few 
$\sim10^3$ K. At $z\approx23$, the first Pop~II star formation event takes place in metal-enriched gas, 
which was previously polluted by a Pop~III star, whose mass of
$20\msun$ was again randomly drawn from the underlying IMF 
Pop~III star. The time delay between the onset of second-generation, Pop~II, star formation and the death of the $20\msun$ Pop~III star 
is $\sim16$~Myr, consistent with our earlier study (\citealp{Jeon2014b}).
By the end of the simulation at $z\approx10.5$, a total of $\rm N_{\rm Pop~II}=65$ Pop~II clusters and $\rm N_{\rm Pop~III}=183$ individual Pop~III stars has formed, 
with 128, 2, 44, and 11 Pop~III stars ending their lives as CCSNe, PISNe, BHs, 
and HMXBs, respectively.
\par
Fig.~\ref{Fig:SFRD} shows the comoving star formation rate density (SFRD) for the Pop~III individual stars and 
the Pop~II clusters. During the initial 50 Myr, only Pop~III stars exist. However, once metal pollution commences 
by Pop~III SNe, the metallicity of the gas increases to trigger the transition 
of the star formation mode from Pop~III to Pop~II. Therefore, at $z\lesssim23$, two star formation modes coexist over the next 
$\sim350$ Myr. As the dense gas available for star formation gets metal-enriched, the 
dominant star formation mode changes from Pop~III to Pop~II. Accordingly, the SFRD for Pop~II stars begins to increase 
and reaches a few $\sim10^{-2} \unit{\msun yr^{-1} Mpc^{-3}}$ at $z\approx14$. 
On the contrary, Pop~III star formation starts to decline from $\sim7\times10^{-3} \unit{\msun yr^{-1} Mpc^{-3}}$ at $z\approx14$ to a few 
$\sim10^{-3} \unit{\msun yr^{-1} Mpc^{-3}}$ at $z\approx10.5$ 
as the amount of dense pristine gas decreases. 
\par
It is noteworthy that the SFRDs measured in the simulation are very episodic rather 
than continuous. This episodic, bursty star formation is mainly caused by a stellar feedback cycle: dense gas inside a 
halo is evacuated by photoheating from stars, the subsequent SN blastwave, or both, rendering it 
gas-starved for a certain time, until fresh gas is provided through smooth accretion or 
mergers with other haloes. Such self-regulated star formation on short timescales is also exhibited by simulations of more massive galaxy formation
\citep{Hopkins2013}, where stellar feedback, i.e. photoheating, SNe, winds, and 
radiation pressure, is taken into account with approximate sub-grid prescriptions. For comparison, we overplot the analytic fitting formula, derived by 
\citet{Hernquist2003}, for higher-mass systems, where atomic cooling is active, which 
agrees well with the SFRD for Pop~II stars predicted by our simulation. We, however, should note that the SFRDs both 
for Pop~III and Pop~II stars are higher than those computed in the adaptive mesh refinement simulation of 
\citet{Wise2012} by a factor of $\sim5$. The reason for this is the biased nature of our zoomed, high-resolution region, focusing on a small volume and omitting under-dense void regions, which results in enhanced star formation activity
compared to the cosmic mean.
\par

\subsubsection{Global impact of stellar feedback}
Fig.~\ref{Fig:snap_large} shows images of the dark matter overdensity, ionized hydrogen fraction, 
gas temperature, and gas metallicity at $z\approx10.5$ in the high-resolution 
region, corresponding to $\sim350$ comoving kpc across, centered on the site of the emerging target galaxy.
The virial radius of the halo hosting the target galaxy, depicted as the white dashed circle in 
Fig.~\ref{Fig:snap_large}, grows over time from $r_{\rm vir}\approx100$ pc at $z=28$ to $r_{\rm vir}\approx1.1$ kpc 
at $z=10.5$. Correspondingly, the halo grows in mass from $5\times10^5\msun$ to $8\times10^7\msun$.
Both the gas inside haloes and in the diffuse IGM are subject to a variety of heating
sources, such as UV radiation from Pop~III and Pop~II stars, X-rays 
from accreting BHs and HMXBs, and thermalized SN explosion energy. 
The temperature inside HII regions reaches $\rm T\simeq 10^4$ K by stellar photoionization heating, and
further increases to $10^7-10^8$ K due to SN shock heating. Such high-temperature 
gas is primarily cooled via inverse Compton cooling off the cosmic microwave background (CMB) and free-free emission to $10^6$ K, and then 
H and He recombination cooling begins to dominate at the lower temperature of $10^4-10^6$~K \citep{Greif2007}.
\par
\begin{figure*}
  \includegraphics[width=125mm]{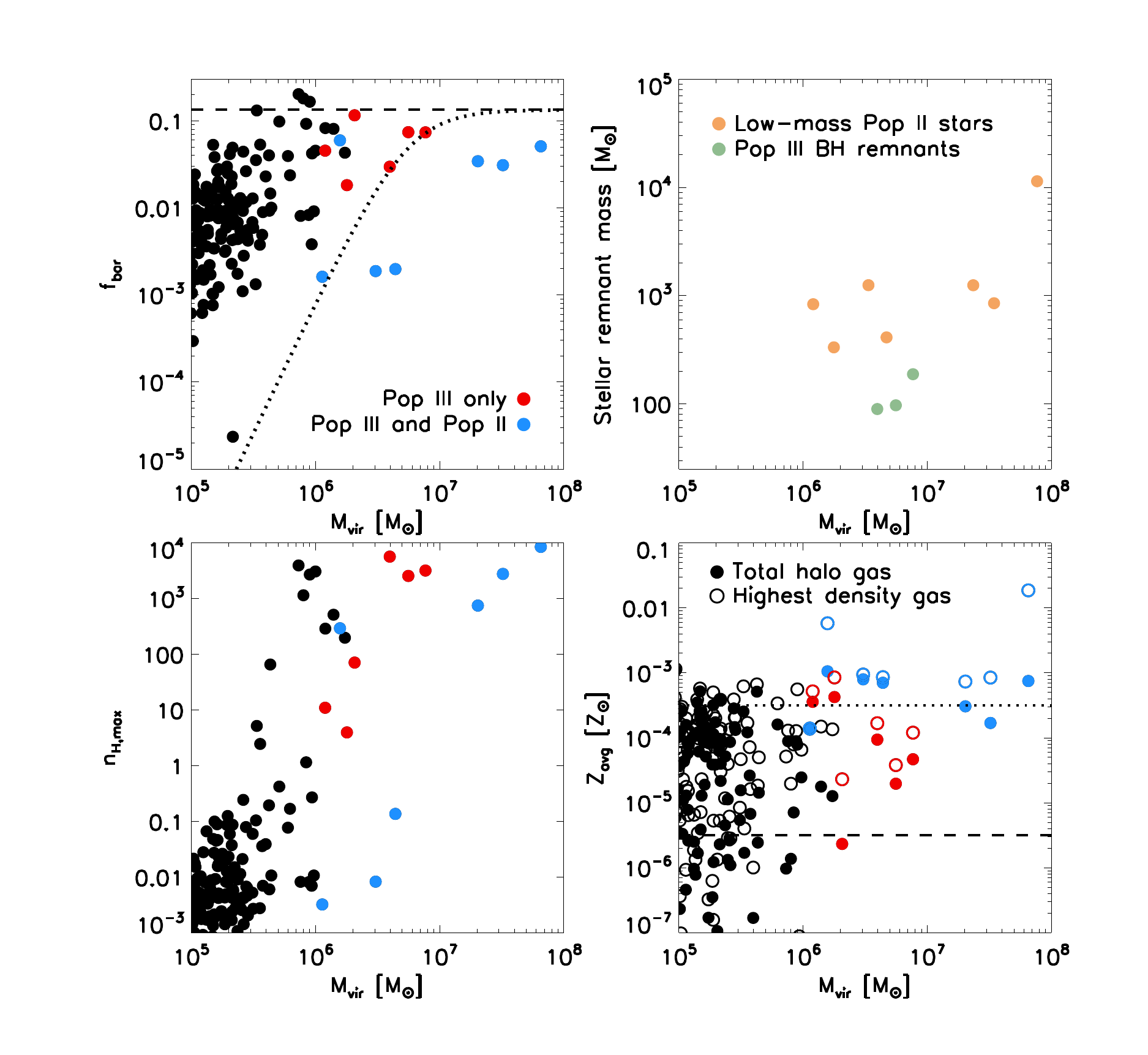}
     \caption{The properties of gas, stars, and metals in haloes at $z\approx10.5$, extracted from the high resolution region 
     using the SUBFIND halo finder (\citealp{Springel2001}). {\it Clockwise from upper left:} halo baryon fraction,
     total mass in Pop~III BH remnants and long-lived (low-mass) Pop~II stars,
     average metallicity of the total halo gas and of the high-density gas, and
the maximum hydrogen number density.
     These properties are computed by mass-weighted averaging inside the virialized 
     region centered on the most bound halo particle. Due to the feedback
     from photoheating, SNe, or both, the gas inside the haloes is significantly evacuated, 
     giving rise to baryon fractions below the cosmic mean value (dashed horizontal line in the upper-left panel). For comparison, we also show the predicted suppression in
baryon fraction due to the presence of a global, external UV background, as given in Equ.~8 (dotted line
in the upper-left panel).
     About $46\%$ of star-forming haloes (colored circles) only host a single 
     or a few Pop~III stars (red circles), whereas the remainder hosts both populations (blue circles). The haloes more massive 
     than $M_{\rm vir}=10^6\msun$ are substantially 
     metal-enriched, giving rise to an average metallicity above $Z=10^{-5.5}\zsun$. The dotted and dashed horizontal lines 
     in the bottom right panel denote the critical metallicity for the Pop~III/Pop~II transition, $Z_{\rm crit}=10^{-3.5}\zsun$ 
     and $Z_{\rm crit, dust}=10^{-5.5}\zsun$, corresponding to metal fine-structure line cooling and dust continuum cooling, 
     respectively. Depending on the choice of critical metallicity, the timing of the Pop~III to Pop~II transition in 
     a given halo can vary. \label{Fig:haloes}}
\end{figure*}

\begin{figure}
  \includegraphics[width=77mm]{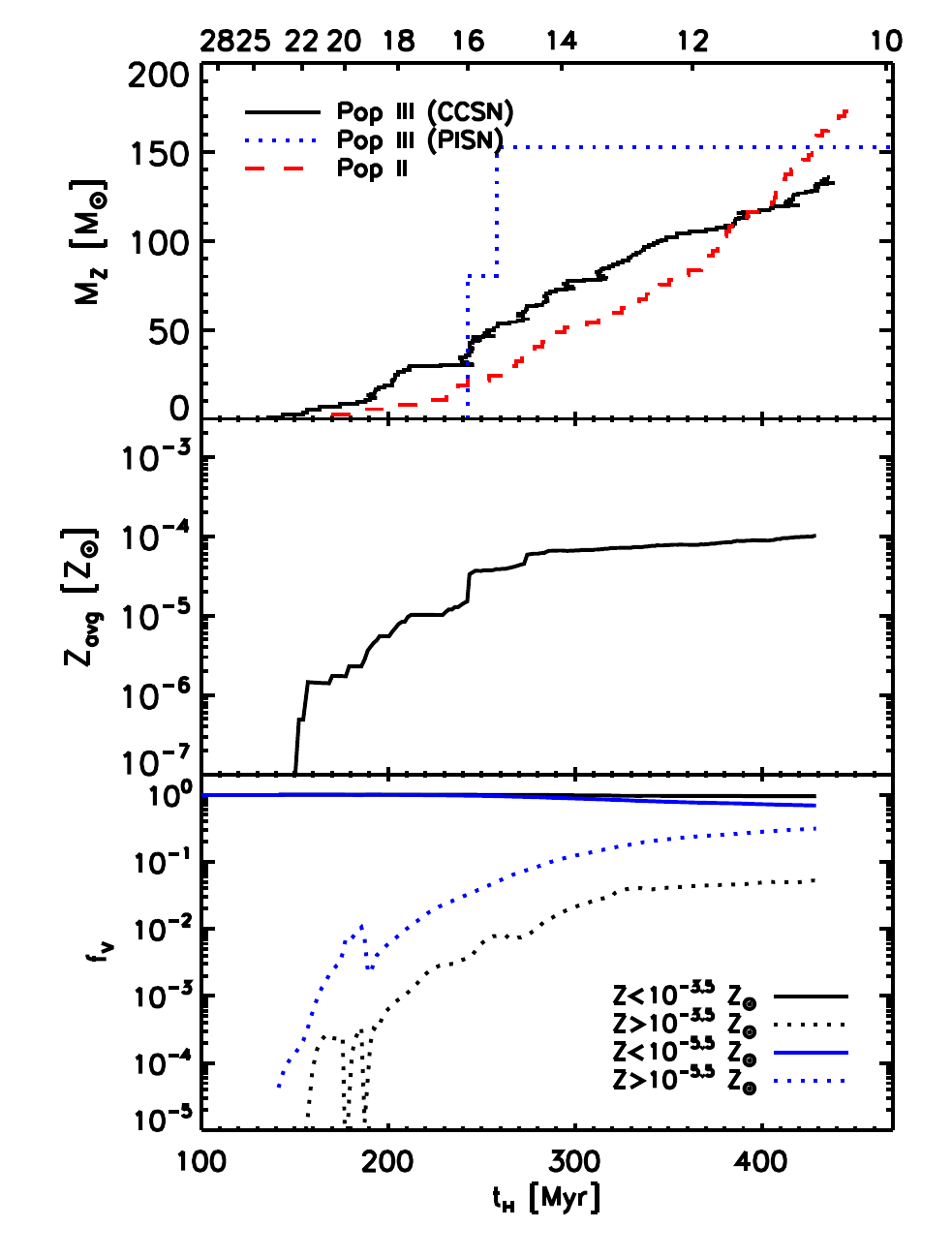}
     \caption{History of the global metal enrichment. As a series of SNe are triggered, the total amount of 
     metals in the simulated region increases over time. Due to the large yield, $y=0.05$, of Pop~III CCSNe, compared to the effective yield of 
     Pop~II SNe, $y_{\rm eff}=0.005$, the former dominate, a situation that is
     reversed at later times, $z\lesssim11$. In terms of the total metal budget, about 75$\%$ of the heavy elements are produced by only two PISNe 
     owing to their extremely large yield, $y\approx0.5$, in combination with their large progenitor mass. The average 
     metallicity increases up to $Z_{\rm avg}=10^{-4} \zsun$ at $z\approx10.5$. On the other hand, the 
     majority of the simulated volume remains almost pristine, below $Z\lesssim10^{-3.5}\zsun$ or $Z\lesssim10^{-5.5}\zsun$,
     as shown in the bottom panel.\label{Fig:Metal_IGM}}
\end{figure}

\begin{figure}
  \includegraphics[width=83mm]{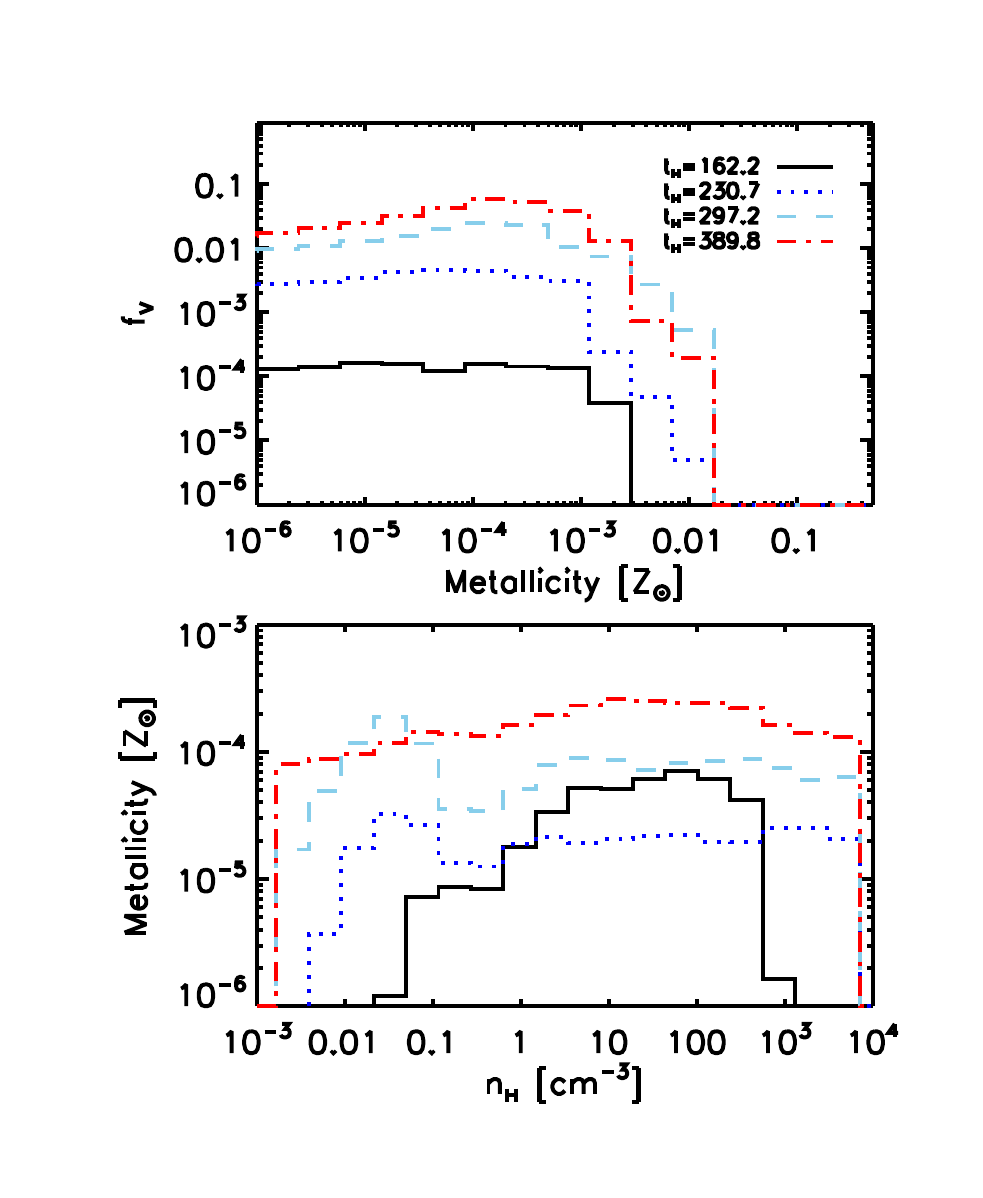}
     \caption{{\it Top:} Evolution of the metal-enriched volume filling fraction as a function of metallicity 
     at four different times. As more and more regions are polluted by SNe, the volume filling fraction increases 
     over time, but the peaks remain at the same metallicity $Z=10^{-4}-10^{-3}\zsun$ at all times. We normalize the 
     volume by the total simulated volume, most of which is still pristine as shown in the 
     bottom panel of Fig.~\ref{Fig:Metal_IGM}. {\it Bottom:} The 
     distribution of metallicity depending on the gas density. Which range of gas density is preferentially 
     metal-enriched is sensitively determined by the gas conditions prior to a SN explosion. For a SN with a low 
     mass progenitor, high density gas ($n_{\rm H}=100\unit{cm^{-3}}$) is likely to be polluted, while 
     the low density gas ($n_{\rm H}=0.1-0.001\unit{cm^{-3}}$) is preferentially enriched by a SN triggered by a massive progenitor.
     \label{Fig:hist_IGM}}
\end{figure}

In order to disentangle the contribution of radiation sources to ionization and heating, in 
Fig.~\ref{Fig:photons} we compare the ionizing photon rates from Pop~III stars, Pop~II stars, and HMXBs. 
While the ionizing photons from Pop~III stars are responsible for ionizing the gas at high redshifts $z\gtrsim13$, 
the contribution from Pop~II stars becomes competitive at later times. On the other hand, the contribution 
from a handful of HMXBs is negligible owing to their rarity and their short lifetime, $\Delta t_{\rm HMXB}=2$~Myr, although the latter quantity is very uncertain and may be
larger.
In \citet{Jeon2014a}, the impact of X-rays from HMXBs on star formation was similarly 
negligible, while the effect of X-ray heating in the diffuse IGM was significant, smoothing out small-scale structures. This smoothing effect is
less prominent in the current simulation, which hosts only a single HMXB by $z\simeq 18$, instead of the 8 sources present at that time in our previous study. 
\par
We assess the impact of the stellar feedback on individual haloes by using the SUBFIND halo finder
(\citealp{Springel2001}) to identify haloes from the highly-refined region in the simulated volume, allowing us to investigate 
their gas, stellar, and metal properties. Each circle in Fig.~\ref{Fig:haloes} marks the mass-weighted average 
of the corresponding stellar or gas properties inside the virial sphere, which is 
centered on the most bound DM particle of 
the halo, and bounded by radii where the DM density is 200 times higher than the mean density of the Universe. 
Star formation takes place only in haloes more massive than $M_{\rm vir}=5\times10^5\msun$, below 
which the haloes are not massive enough to satisfy the condition 
$t_{\rm cool} < t_{\rm ff}$, where $t_{\rm cool}$ and $t_{\rm ff}$ are the cooling time and the free-fall time, 
respectively, for the gas to cool and form stars (e.g. \citealp{Tegmark1997}; \citealp{Bromm2002}).
\par
Fig.~\ref{Fig:haloes} shows the gas properties of the extracted haloes at $z\approx10.5$, including the 
baryon fraction, the total mass of Pop~III BH remnants and Pop~II low-mass (long-lived) stars, the highest gas density achieved, and the average metallicity of the gas in the
halo. Photoheating and SNe drive the gas out of the haloes, lowering the baryon fraction 
below the cosmic mean (upper-left panel of Fig.~\ref{Fig:haloes}). 
Among the haloes more massive than $10^6\msun$ at $z\approx10.5$, about $40\%$, 
corresponding to 7 out of 17 haloes (blue circles in Fig.~\ref{Fig:haloes}), experience 
photoheating and SN blastwave heating by both Pop~III stars and Pop~II clusters. 
Specifically, the three most massive haloes, ($M_{\rm vir}\sim7\times10^7\msun$, $\sim3\times10^7\msun$, $\sim2\times10^7\msun$) 
at $z\approx10.5$ have experienced (32, 9, 7) Pop~III SNe and (28, 2, 3) SNe 
from Pop~II clusters. The gas fractions of these haloes at $z\approx10.5$ are $5.1\%$, $3.3\%$, and $3\%$, 
respectively. For lower-mass haloes, on the other hand, the scatter in the gas fraction is much more substantial, due to their susceptibility
to negative stellar feedback in the shallow potential wells. Those haloes are more likely to lose their gas
once they undergo star formation, and it takes longer for the ejected gas to be re-incorporated.
\par
About $35\%$ of the haloes larger than $M_{\rm vir}\gtrsim10^6\msun$, denoted as red circles, only host
Pop~III stars. Most of them experience subsequent CCSN explosions, whereas 2 haloes host a rare PISN.
The effect of a PISN is much more disruptive than that of a CCSN. As an example,  
the third heaviest halo hosts a $160\msun$ PISN at $z=16$, when its mass reaches 
$M_{\rm vir}\sim10^7\msun$. This hyper-energetic explosion effectively evacuates the gas within the halo, lowering the baryon fraction by an order of magnitude, from 
$f_{\rm bar}=0.088$ at $z=16$ to $f_{\rm bar}=0.0046$ at $z=13.7$. Subsequently, 
the halo gas begins to be replenished through accretion of both the ejected metal-enriched gas and 
fresh primordial gas along the cosmic filaments, such that at the end of the simulation the baryon fraction again reaches $f_{\rm bar}=0.03$. 
Despite the gas recovery, however, any further star formation inside the halo has been completely suppressed since 
$z=16$. 
\par
Haloes with $M_{\rm vir}\lesssim10^8\msun$, corresponding to the mass range considered in this work, are susceptible not only to the local 
stellar feedback inside the host itself, but also to the external global UV background that also photoheats the gas in their shallow 
potential wells, reducing the baryon fraction. For comparison, we overplot a fit (dotted line the upper-left panel of Fig.~\ref{Fig:haloes}) that 
describes how the baryon fraction is affected by reionization as a function of halo mass and redshift 
(\citealp{Gnedin2000}; \citealp{Okamoto2008}):
\begin{equation}
f_{\rm bar}(M,z) = <f_{\rm bar}>\left( 1+ (2^{\beta/3}-1) \left[ \frac{M}{M_{\rm c}(z)}\right]^{-\beta} \right)^{-3/\beta},
\end{equation}
where <$f_{\rm bar}$> is the cosmic mean baryon fraction, $\beta=2$ a fitting parameter, and
$M_{\rm c}(z=10)=7\times10^6\msun$ the characteristic mass, below which galaxies are strongly affected by the UV background. 
The gas in haloes affected by internal feedback from both Pop~III and Pop~II stars tends to be more strongly evacuated, compared to the external UV background case, whereas
$f_{\rm bar}$ in haloes hosting only Pop~III stars is less severely attenuated.
\par
The average gas metallicity in the haloes with $M_{\rm vir}\gtrsim10^{6}\msun$ lies between 
$Z_{\rm avg}=10^{-5}-10^{-3}\zsun$. The metal-enrichment in the individual haloes is mainly achieved 
by CCSNe, because the relatively low characteristic mass, $M_{\rm char}=20\msun$, for the Pop~III IMF 
results in only two PISNe during the run. When the $160\msun$ PISN is triggered, the surrounding gas 
is immediately polluted to an average of $Z=4\times10^{-3}\zsun$. However, those 
heavy elements are quickly driven out beyond the virial radius 
of the host halo. After $50$ Myr, the ejected metals start to fall back as gas is accreted, and the average gas metallicity 
reaches $Z=3\times10^{-4}\zsun$ in the halo with a mass of 
$M_{\rm vir}\approx2\times10^7\msun$ at $z\approx10.5$. This agrees with the results by 
\citet{Wise2012}, where they found that for haloes $M_{\rm vir}>3\times10^7\msun$, all haloes are 
enriched above $\rm [Z/H]>-4$. However, we should point out that the metal contributors were Pop~III PISNe 
in their work, while the haloes at $z=10.5$ in the current work are mainly polluted by CCSNe. At $z\approx10.5$, if we consider 
all haloes with a mass above $M_{\rm vir}=10^5\msun$, only $2\%$ of them experience in-situ star formation, 
while 25$\%$ of the haloes exhibit an average gas metallicity above $Z=10^{-6}\zsun$. 
Thus, some of them were enriched by SNe in nearby haloes.
\par
Given that star formation takes place in cold dense gas, specifically considering the metallicity of the densest gas 
in the haloes is meaningful. In the most massive three haloes, 
the densest gas is enriched above $Z_{\rm crit}=10^{-3.5}\zsun$ (see the bottom-right panel of Fig.~\ref{Fig:haloes}), 
indicating that Pop~II stars are expected to form out of the gas. However, it does not necessarily imply that the haloes cease
forming Pop~III stars. Instead, Pop~III stars may
continue to form until the end of the simulation in the highest density regions, which are already metal-enriched, but possibly insufficiently so
to enable Pop~II star formation. The time at which the Pop~III/Pop~II transition occurs depends on the 
choice of the critical metallicity, which is still quite uncertain. For instance, the densest clouds (open circles above $10^3$\unit{cm^{-3}} 
in the bottom-left panel of Fig.~\ref{Fig:haloes}) in the haloes between 
$3\times10^6\msun\lesssim M_{\rm vir}\lesssim10^7\msun$ are about to form stars. Based on 
$Z_{\rm crit}=10^{-3.5}\zsun$, which is assumed in this work, some Pop~III stars are expected to form, 
whereas Pop~II stars would form if we adopted a critical metallicity, $Z_{\rm crit}=10^{-5.5}\zsun$ (dashed horizontal line in the bottom-right panel of Fig.~\ref{Fig:haloes}), 
set by dust-continuum cooling (e.g. \citealp{Omukai2000}; \citealp{Bromm2001b};
\citealp{Schneider2002}; \citealp{BrommLoeb2003};
\citealp{Omukai2005}; \citealp{Schneider2010}; \citealp{Schneider2012};
 \citealp{Chiaki2014}; \citealp{Ji2014}). Therefore, the latter case would give
rise to an earlier transition in star formation, and would more
rapidly extinguish the Pop~III mode.
\par
\begin{figure*}
  \includegraphics[width=130mm]{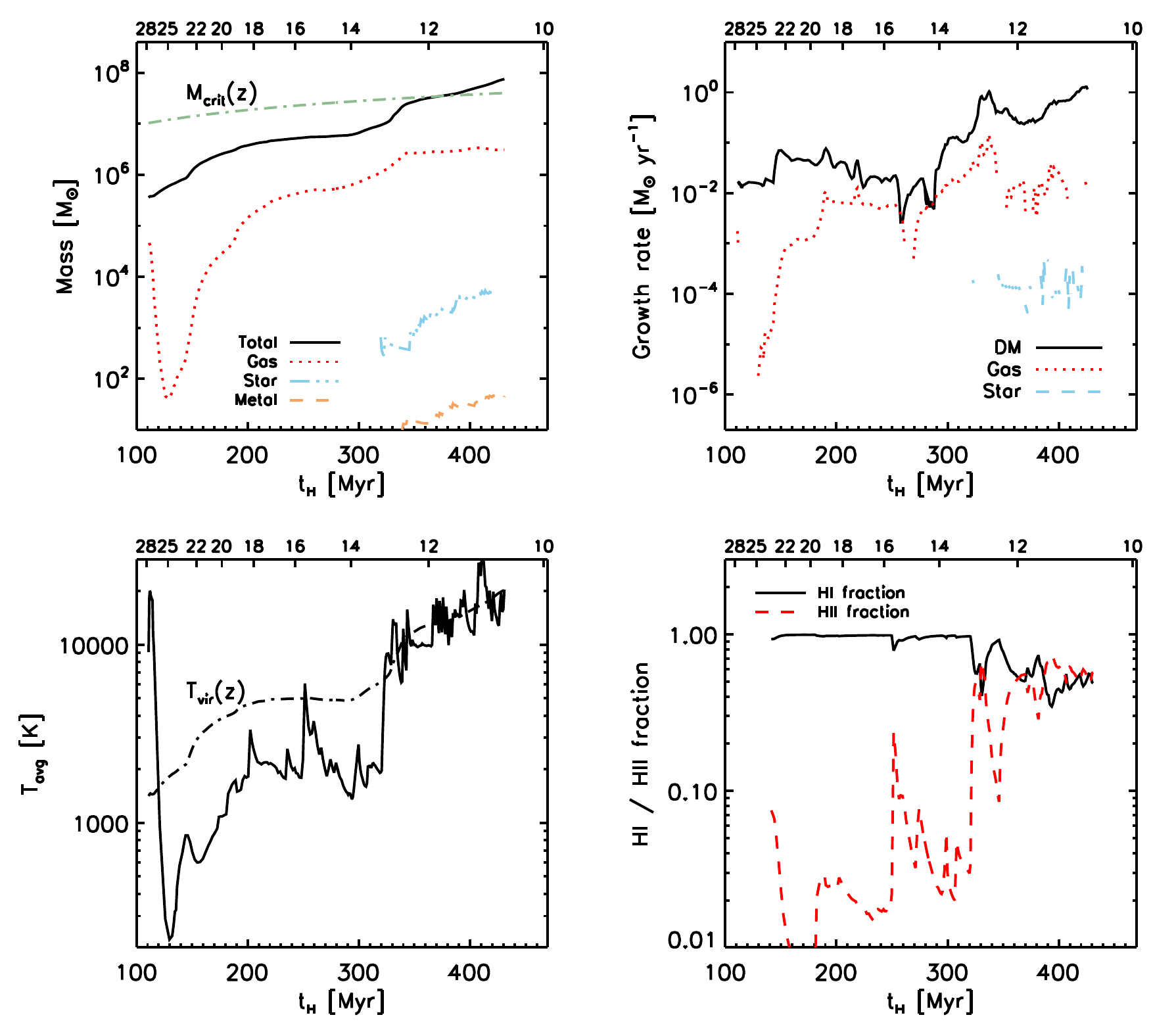}
     \caption{Assembly history of the emerging target galaxy. {\it Clockwise from upper left}: 
     Mass growth of the target galaxy, growth rate of various halo components, 
     H~I/H~II fractions, and the average gas temperature. Starting from 
     $M_{\rm vir}\approx5\times10^5\msun$ at $z\approx28$, the galaxy grows in mass and reaches 
     the threshold for atomic hydrogen cooling at $z\approx11.7$. The final virial mass 
     at the end of the simulation, $z\approx10.5$, is $M_{\rm vir}\sim10^8\msun$. The stellar content of the galaxy is dominated 
     by Pop~II, with a total mass of $M_{\ast}\approx6\times10^3\msun$, formed
with an effective star formation rate of
     $\sim10^{-4}\msun\unit{yr^{-1}}$. Through a complex interplay 
     between metal ejection and re-accretion along the filaments of the cosmic web, roughly $\sim40\msun$ in metals 
     are present inside the galaxy at the end of the simulation.
  \label{Fig:fg_evol}}
\end{figure*}
\subsubsection{Metallicity evolution}
By the end of the simulation, a total of 128 out of 183 Pop~III stars end up as CCSNe 
and 2 stars explode as PISNe. Fig.~\ref{Fig:Metal_IGM} displays the resulting metal enrichment history of the 
simulated region. While the gas confined inside the haloes is mainly enriched by Pop~III and Pop~II CCSNe, the enrichment in the diffuse IGM is dominated by the rare PISN events. Indeed, although
only two PISNe have occurred, they produce a total of $\sim150\msun$ in metals, accounting 
for $75\%$ of the total metal budget at $z\approx10.5$. Intriguingly, the optimal strategy to search for PISN-enriched material may thus be to probe the high-redshift IGM with absorption spectroscopy \citep{Wang2012}. The average mass-weighted metallicity of the region increases to $10^{-4}\zsun$ 
at $z\approx10.5$. However, the majority of the region is metal-enriched below the critical metallicity $Z_{\rm crit}=10^{-3.5}\zsun$, 
giving rise to a volume filling factor of $f_V\approx0.95$ $(\lesssim 10^{-3.5}\zsun)$ and 
$f_V\approx0.69$ $(\lesssim 10^{-5.5}\zsun)$. 
The volume filling fraction, $f_V$, is defined as 
\begin{equation}
f_V = \frac{\sum_{\rm subset}{m_i / \rho_i}}{\sum_{\rm total}{m_i / \rho_i}}.
\end{equation}
For gas that is sufficiently metal-enriched to form Pop~II stars, the volume filling fraction 
keeps increasing and reaches $f_V\approx0.05$ $(\gtrsim10^{-3.5}\zsun)$ and $f_V\approx0.29$ $(\gtrsim10^{-5.5}\zsun)$ 
at $z\approx10.5$.
\par
The upper panel of Fig.~\ref{Fig:hist_IGM} shows the volume filling fraction of the gas as a function of metallicity at four 
different characteristic times. Overall, the metal-enriched volume fraction increases over time, peaking at 
$Z\approx2-3\times10^{-4}\zsun$ and declining at high 
metallicities $Z>10^{-3}\zsun$. We also show the average 
metallicity as a function of gas density at the four selected times. The metallicity distribution 
sensitively depends on the gas density at the time when the SN is about to explode. The initial conditions for a SN 
explosion are set by the photoheating from its progenitor star. If the stellar radiative feedback is strong 
enough to evacuate the gas around the star, lowering the gas density to $n_{\rm H}=0.1-10^{-3}\unit{cm^{-3}}$, 
this low density gas is first likely to be polluted. For lower mass progenitor stars, on the other hand, due to their weaker ability 
of evacuating gas in their surroundings, such high density gas is preferentially enriched by the explosion.
\par
This trend is clearly displayed in the bottom panel of Fig.~\ref{Fig:hist_IGM}. The progenitor mass of the first SN, 
triggered at $t_{\rm H}=133$ Myr, is $25\msun$. The outflow, driven by the photoheating of the 
$25\msun$ Pop~III star, is not strong enough to evacuate the surrounding dense gas. Therefore, the subsequent SN ejecta encounter this high density gas first, 
giving rise to the metallicity peak at $n_{\rm H}=100 \unit{cm^{-3}}$ 
(see solid line in the bottom panel of Fig.~\ref{Fig:hist_IGM}). On the contrary, sufficiently strong radiative 
feedback from massive progenitor stars sets the stage for the following SN explosions to occur in low-density gas, with
$n_{\rm H}=0.01-0.1\unit{cm^{-3}}$. This low density gas is then preferentially enriched, corresponding 
to the cases shown as dotted and dashed lines in Fig.~\ref{Fig:hist_IGM}. In summary, the derived trend is as follows: a SN with 
a low mass Pop~III progenitor tends to pollute dense gas while a massive progenitor followed by a SN is more likely to 
enrich low density gas. This trend, however, disappears as both low and high density 
regions are equally polluted as SN events become more frequent.

\begin{figure*}
  \includegraphics[width=130mm]{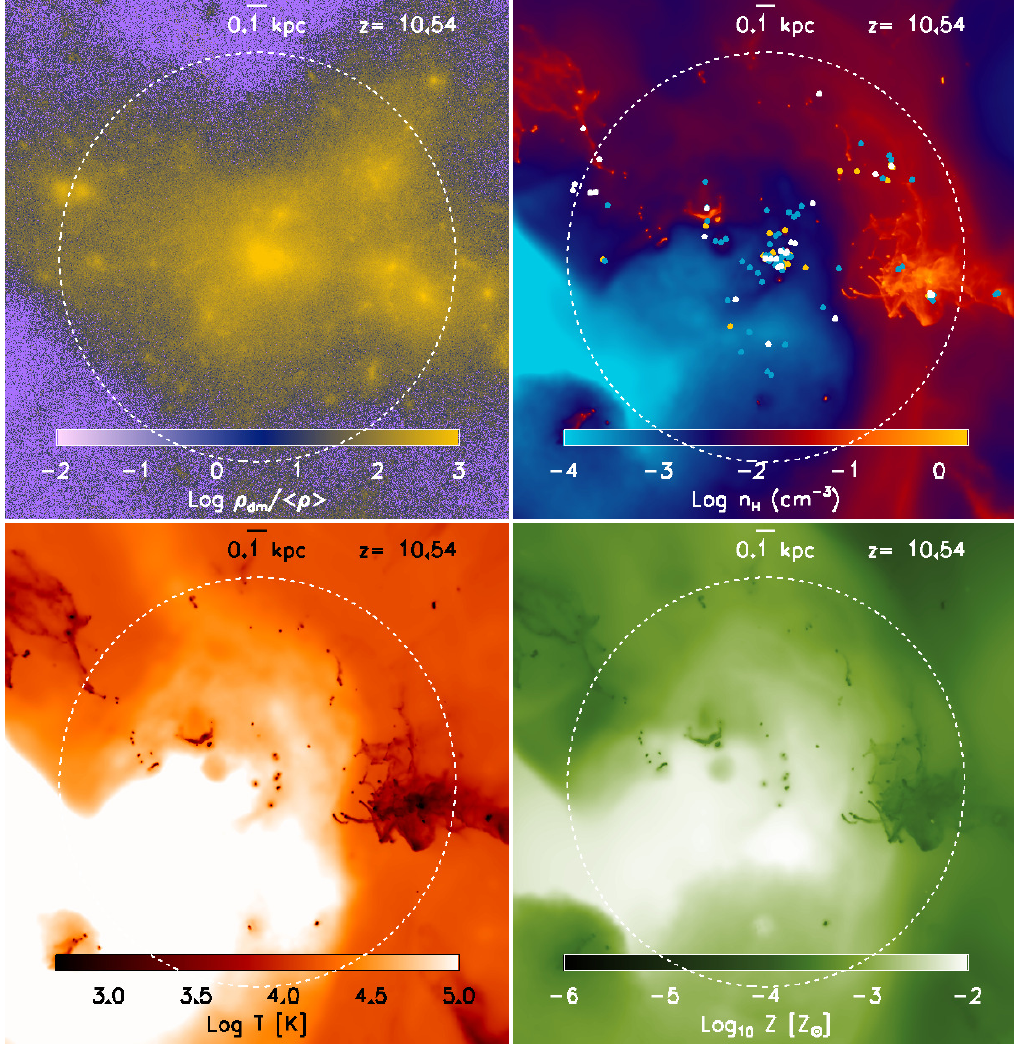}
     \caption{Morphology in the target galaxy at $z\approx10.5$. {\it Clockwise from upper left}: Dark matter overdensity, 
     hydrogen number density, gas metallicity, and gas temperature, averaged along the line of sight within 
     the central $\simeq80$ kpc (comoving). Dashed white circles denote the virial 
     radius, which is $r_{\rm vir}\approx1.1$ kpc at this time, with a corresponding virial mass of 
     $M_{\rm vir} \approx 8\times10^7\msun$. The positions of the Pop~III remnants 
     are marked as filled circles in the upper-right panel: BHs as yellow, neutron stars left behind by lower-mass CCSNe as blue, whereas the location of Pop~II clusters are 
shown in white.
     Due to the ongoing star 
     formation inside the galaxy, the central gas is substantially evacuated into the void that is roughly 
     perpendicular to the filaments of the cosmic web, resulting in gas densities $n_{\rm H}\lesssim10^{-4}\unit{cm^{-3}}$ with temperatures  
     well above $10^6$~K due to SN blastwaves. The central region is substantially 
     metal-enriched to $Z\approx10^{-2}\zsun$, and all gas inside the virial radius is polluted above 
     $Z=10^{-5}\zsun$. \label{Fig:snap_small}}
\end{figure*}

\begin{figure}
  \includegraphics[width=85mm]{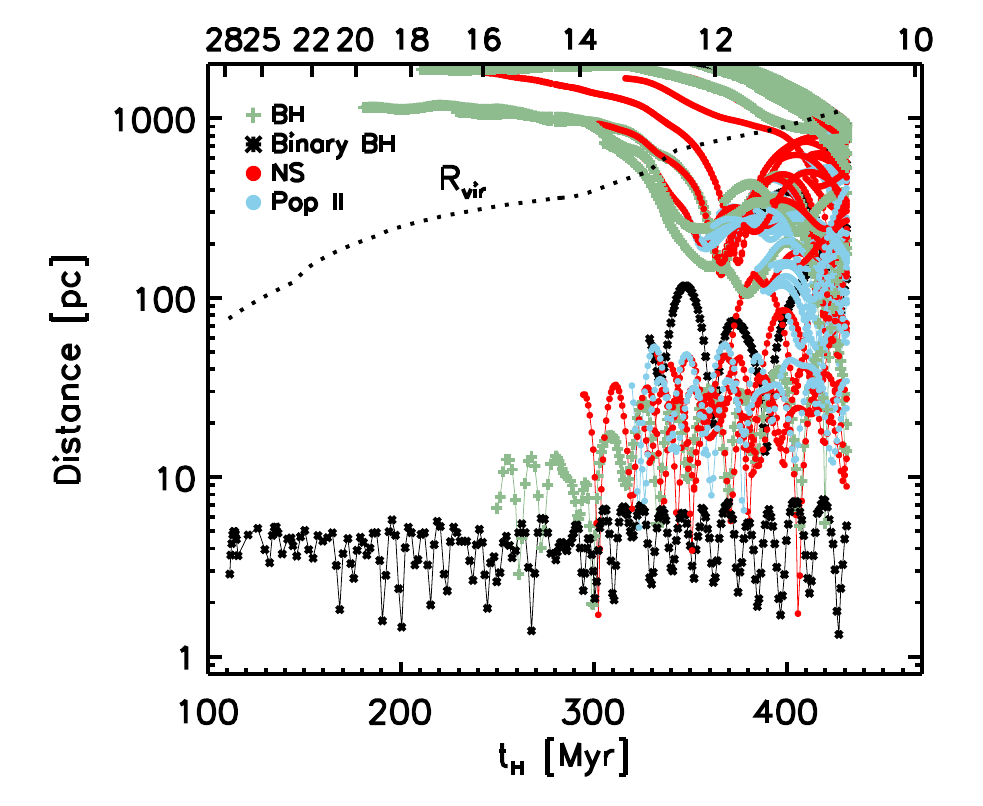}
     \caption{Trajectories of all stellar remnants formed during the assembly process of the target galaxy.  
     The dotted line denotes the virial radius of the halo. As can be seen, about $30\%$ of the remnants, 
     corresponding to $15\%$ of total remnant mass, formed in neighboring haloes and were incorporated 
     into the target halo. The majority of Pop~II clusters are born in-situ, out of gas inside the galaxy, and only 3 clusters arise
     from mergers. \label{Fig:fg_star}}
\end{figure}

\begin{figure}
  \includegraphics[width=65mm]{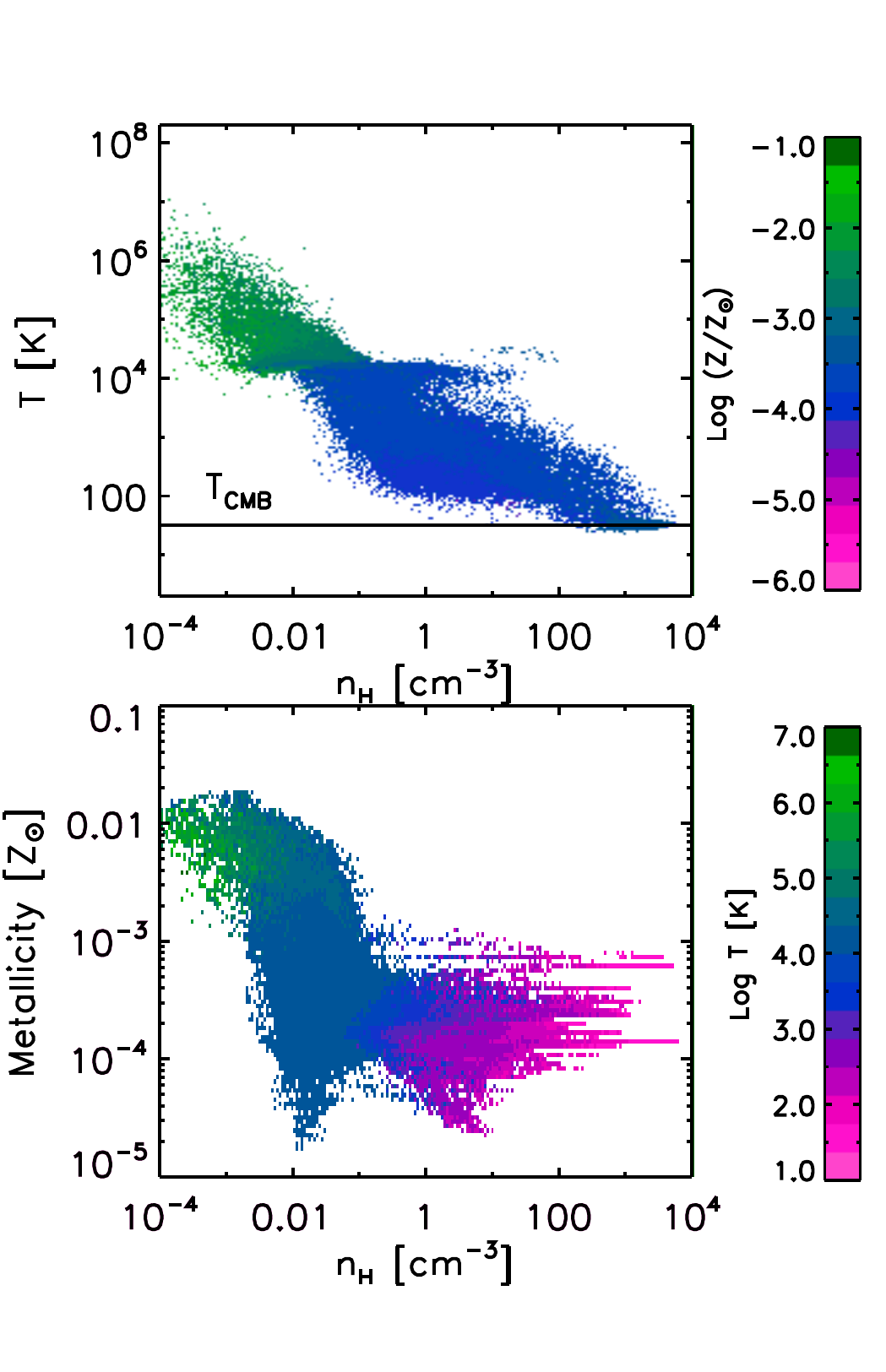}
     \caption{ {\it Top}: Density-temperature phase diagram of the gas 
     inside the target galaxy at $z\approx10.5$. The color coding indicates gas metallicity. 
     Thanks to the metal cooling, high density gas ($n_{\rm H}\gtrsim 100 \unit{cm^{-3}}$) 
     is able to cool down to the CMB temperature floor at $\sim30$ K. 
     {\it Bottom}: Density-metallicity phase diagram. The color coding now indicates gas temperature. 
     The metallicity spread in the range $Z=10^{-4}-10^{-3}\zsun$ for high density gas, 
     which is mainly responsible for subsequent star formation, implies that both
     Pop~III and Pop~II might form in the enriched gas, but this depends on the choice of critical metallicity (see main text). Note that 
     the gas with $n_{\rm H}<0.01\unit{cm^{-3}}$ is highly enriched to $\sim0.01\zsun$, 
     due to the recent SN explosion in the rarified region established by photoheating from its progenitor.
     \label{Fig:fg_metal}}
\end{figure}

\begin{figure}
  \includegraphics[width=90mm]{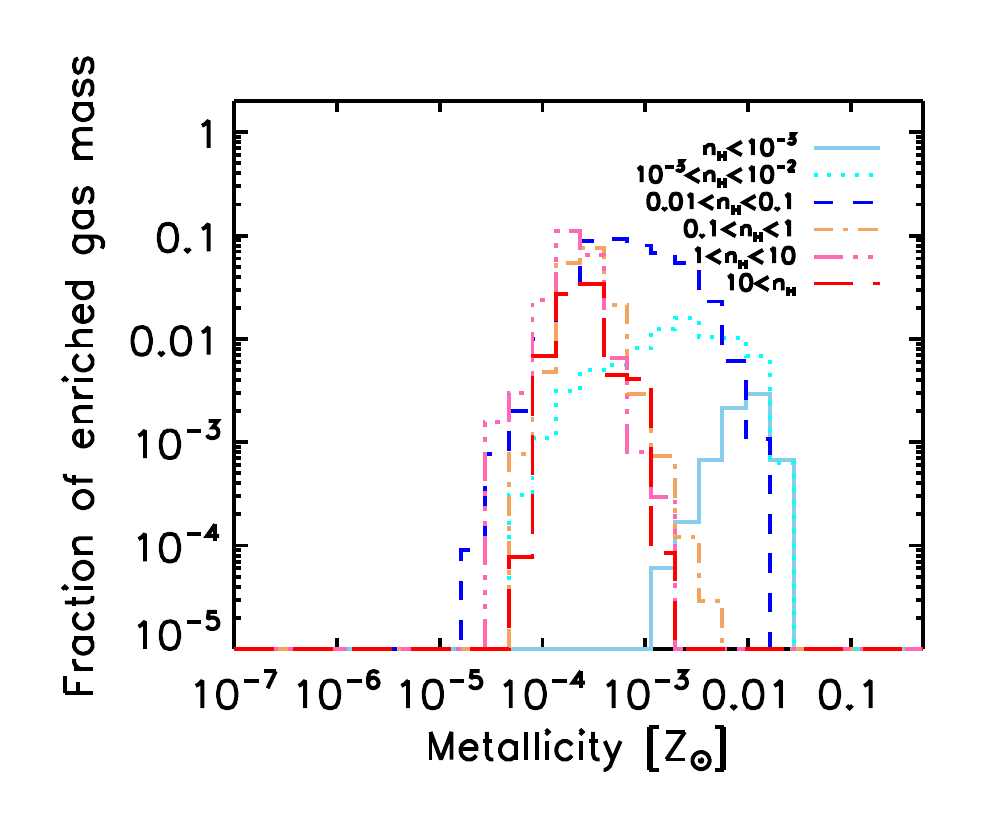}
     \caption{Enriched gas mass fraction within the virial radius of the galaxy as a function 
     of metallicity at $z\approx10.5$. It can be seen that the peak remains 
     roughly at $Z=10^{-4}-10^{-3}\zsun$ for all densities, except for the lowest density gas, with $n_{\rm H}<0.01\unit{cm^{-3}}$, which 
     shows a peak at $\gtrsim10^{-3}\zsun$. This is due to a recent SN explosion, which is preferentially propagating along 
     the underdense region of the galaxy. \label{Fig:fg_metal_dist}}
\end{figure}

\subsection{Assembly of the first galaxy}
\subsubsection{Mass evolution}
The first star in the Lagrangian volume of the emerging target galaxy forms at $z\approx28$ out of 
the primordial gas within the main progenitor minihalo of mass $\sim5\times10^5\msun$. After its short lifetime 
of 2.7~Myr, the star collapses into a BH, acting as a HMXB source for the next 2~Myr. Stellar and BH feedback 
evacuate the gas by reducing the baryon fraction of the halo from $f_{\rm bar}=0.12$ at $z\approx28$ to essentially zero at $z\approx25$. Afterwards, the baryon fraction again increases, and at $z\approx16$ gas densities have increased to $n_{\rm H}=10^4 \cmci$, thus enabling further star formation. Subsequently, the baryon fraction remains above 
$f_{\rm bar}=0.05$ throughout since $z\lesssim16$, even in the presence of ongoing star formation, as the halo is sufficiently massive, $M_{\rm vir}\gtrsim5\times10^6\msun$, to hold onto 
its gas despite feedback.
\par
The first Pop~II stars in the target halo form at $z=13.1$ out of
gas that was enriched by a $33\msun$ Pop~III star that ended its 
live as a CCSN, thus initiating the prolonged chemical enrichment process. Until $z=13.3$, the halo grows in mass 
mainly via smooth accretion along the filaments of the cosmic web and through minor mergers. At $z\approx13.3$, 
the halo experiences a major merger and doubles its mass to $2\times10^7\msun$ at $z\approx12.6$, 
over a time span of $\sim25$~Myr. The average growth rate 
during this period is $\dot{M}\approx0.5\msun\unit{yr^{-1}}$. At $z=11.7$, the halo exceeds $3.5\times10^7\msun$, and
fulfills the criterion for the onset of atomic hydrogen cooling (see Fig.~\ref{Fig:fg_evol}). At the end of the simulation at $z\approx10.5$, the target halo has reached a virial mass of
$\sim8\times10^7\msun$, after undergoing another major merger at $z=11.3$.
\par
Fig.~\ref{Fig:fg_evol} also shows the growth rates onto the target halo, the evolution of the average 
gas temperature, and the H~I/H~II fraction. The gas is photoheated by 
Pop~III and Pop~II stars to temperatures of $\sim10^4$ K, temporarily reaching even higher temperatures in the wake of SN explosions 
or due to X-ray heating from accreting BHs and HMXBs. Early on, 
the gas can quickly recombine once an emission source is turned off. However, 
at $\lesssim z\approx12$, about $50\%$ of the gas remains ionized by continuous star formation activity. 
The morphology of the emerging galaxy at $z\approx10.5$ is shown in Fig.~\ref{Fig:snap_small}. Along a direction 
roughly perpendicular to the filaments of the cosmic web, the gas is 
highly evacuated to $n_{\rm H}<10^{-4}$\unit{cm^{-3}} by Pop~II stellar feedback, heating it to $T\sim10^6-10^7$ K 
and enriching it to $Z\sim 10^{-2.5}\zsun$ (see the bottom panels of Fig.~\ref{Fig:fg_evol}). 
\subsubsection{Star formation history}
Until $z=12.5$, star formation in the target halo is dominated by Pop~III stars, but afterwards Pop~II stars form out of 
metal-enriched gas, and become the dominant stellar population. By the end of the simulation, 
the emerging target halo has hosted 15 accreting BHs, 2 HMXBs, 40 Pop~III CCSNe, possibly leaving neutron stars behind, and 31 
Pop~II clusters. The total mass of the stellar remnants increases 
from $\sim100\msun$ at $z\approx28$ to $3\times10^4\msun$ at $z\approx10.5$ (see the left-upper panel of 
Fig.~\ref{Fig:fg_evol}). However, not all of them are formed inside the halo. In order to determine their birth 
places, we trace their trajectories in Fig.~\ref{Fig:fg_star}, by recording their distances 
from the center of the target halo. We find that 30$\%$ of the remnants, corresponding to $15\%$ in mass, formed 
in neighboring haloes and were subsequently accreted by the target halo.
\par
One of the key physical quantities of the galaxy is the total stellar mass, needed to predict its observational signature.
Given the extremely short lifetime of Pop~III stars, of order a few Myr, we only consider Pop~II clusters to estimate the stellar 
mass, computed as
$M_{\rm \ast, Pop~II}\propto \int^{m_{\rm max}}_{m_{\rm min}} m^{-\alpha}m\hspace{0.1cm}dm$, 
where the IMF slope is $\alpha=1.35$, and the lower mass limit $m_{\rm min}=0.1\msun$.
The upper mass limit, $m_{\rm max}$, on the other hand, is time variable and depends on the age of the Pop~II cluster, such that only stars with a main-sequence lifetime larger
than the age of the cluster contribute to the persistent stellar mass. The estimated stellar mass in the halo starts 
from $540\msun$ at $z\approx13$ and increases to $6\times10^3\msun$ at $z\approx10.5$. The corresponding 
star formation rate (SFR) is a few $\sim10^{-4}\msun\unit{yr^{-1}}$.
\par
 The small characteristic mass for Pop~III stars, $M_{\rm char}=20\msun$, adopted
in this simulation, results in a qualitatively very different star formation and galaxy assembly history, compared with the
$M_{\rm char}=100\msun$, which was often used in previous first galaxy formation work \citep{Bromm2011}. For instance, our simulation employs the same initial 
conditions as those in \citet{Greif2010} and \citet{Jeon2012}, both of which used $M_{\rm char}=100\msun$ 
for Pop~III stars and neglected the Pop~III/Pop~II transition. \citet{Greif2010} 
showed that the strong feedback from a single PISN and its progenitor, formed at the 
beginning of their simulation at $z\approx28$, evacuates most of the gas out of the host halo. Except for one additional
star at $z\approx14$, the halo remained without further star formation for 
the next $\sim300$ Myr after the PISN. More stars formed in the simulation of \citet{Jeon2012}, 
where the target halo hosted a total of 8 stars in the presence of the positive X-ray feedback from an accreting BH source.
To the contrary, the same target halo here forms a total of 52 Pop~III stars and 
31 Pop~II clusters by the end of the simulation. This striking difference in the star formation history reflects
the relative strength of radiative and mechanical feedback from Pop~III stars, depending on their masses: 
photoheating from massive Pop~III stars strongly evacuates the gas and 
suppresses further star formation, as demonstrated in \citet{Greif2010} and \citet{Jeon2012}.

\subsubsection{Metal enrichment}
The 40 Pop~III and 248 Pop~II CCSNe that exploded in the target halo produce a total of $28\msun$ 
and $70\msun$ in metals, respectively. Most ($60\%$) of 
the metals are permanently ejected from the target system into the diffuse IGM, leaving 
only $40\msun$ of metals in the galaxy, supplied partially by the re-accretion of metals, or provided by in-situ SN events.
The mass-weighted, average gas metallicity in the galaxy is <$Z$>$=7\times10^{-4}\zsun$ 
at $z\approx10.5$, whereas that of the densest gas, out of which new stars form, is higher by a factor of $\sim$ 30.
\par
A notable difference from many previous studies is that {\it all} the heavy chemical elements inside the target galaxy are produced by CCSNe, and 
not in PISN events. Despite the smaller yield $y=0.05$ for CCSNe, in contrast to $y=0.5$ for PISNe, and the 
smaller masses of $M_{\rm char}=20\msun$ for Pop~III stars, the total mass in metals, now provided by a large number of Pop~III CCSNe, is comparable 
to the metal production by a single $200\msun$ PISN. The result is a level of overall metal enrichment similar to previous studies.
For instance, the central gas within a $M_{\rm vir}\approx10^8\msun$ halo
at $z\approx10$ is enriched to an average of $\sim10^{-3}\zsun$ by a single $200\msun$ PISN (e.g. \citealp{Greif2010}).
With a series of PISN events, the central gas cloud in $M_{\rm vir}=5\times10^7-10^8\msun$ haloes could be 
enriched to $\rm [Z/H]=-2.5-3.0$, as well (e.g. \citealp{Wise2012}; \citealp{Biffi2013}). 
Therefore, we emphasize that while the choice of characteristic Pop~III mass results 
in substantially different star formation histories, as discussed above, the degree of metal enrichment might be 
similar whether a first galaxy is enriched by a single or a few energetic PISNe, or by numerous CCSNe.
\par
To gain further insight into the stellar population mix in our target galaxy, in Fig.~\ref{Fig:fg_metal} we show the gas temperature 
(top) and gas metallicities (bottom) inside the galaxy as a function of hydrogen density at $z\approx10.5$. The metallicities 
of the high-density gas ($n_{\rm H}\gtrsim10^3 \unit{cm^{-3}}$), which is the main reservoir for star formation, are in the range 
$10^{-4}\zsun$ to $10^{-3}\zsun$. As a result, instead of having a clear termination of 
Pop~III star formation, both populations may coexist within the galaxy. Lowering the critical metallicity, however, as suggested by the dust-cooling theories, would lead to exclusively Pop~II star formation inside the target system. 
Gas temperatures reach the floor set by the CMB at $\sim30$ K (horizontal line in the top panel of Fig.~\ref{Fig:fg_metal}), due to efficient metal fine-structure line 
cooling. Without it, the gas would only be able to cool to $\sim200$~K via molecular 
hydrogen cooling.
Overall, $40\%$ of the gas mass within the galaxy is enriched above the critical metallicity for fine-structure cooling, and no primordial gas 
($Z\lesssim10^{-5}\zsun$) remains. As shown in Fig.~\ref{Fig:fg_metal_dist}, across all densities, the gas metallicity peaks at $10^{-4}-10^{-3}\zsun$, 
excluding the lowest densities, $n_{\rm H} \lesssim 0.01\unit{cm^{-3}}$, where the peak is located above $10^{-3}\zsun$. This difference is associated with recent metal pollution in low density regions, which are already highly 
evacuated by photoheating from the massive SN progenitor.
\par
The bright afterglow spectra of possible Pop~III gamma-ray bursts (GRBs), with absorption lines imprinted by intervening metal-enriched clouds, would be an ideal tool for probing the metal enrichment of the early Universe simulated here (e.g. \citealp{Wang2012}; \citealp{Macpherson2013}). Specifically, the equivalent widths of the metal absorption spectra may allow us to distinguish 
whether the first heavy elements were produced by a PISN or a CCSN, and in turn constrain the Pop~III IMF (e.g. \citealp{Wang2012}). As we have discussed in Section~3.1, the diffuse, low column density, IGM may be the preferred hunting ground for a possible PISN enrichment signature, whereas the higher-column density systems may be dominated by the conventional CCSN signature.

\subsection{Stellar and galactic archeology}
 A promising method to constrain the properties of primordial stars is to probe 
the metal-poor stars in the halo of our Milky Way, or in nearby dwarf satellite galaxies. 
This approach, often termed stellar or galactic archeology (e.g. \citealp{Frebel2010}; \citealp{Brown2012}; 
\citealp{Karlsson2013}; \citealp{Vargas2013}; \citealp{Milos2014}), is based on the assumption that such 
metal-poor halo stars were born out of gas that was polluted by a single 
or a few Pop~III SNe, and have survived until the present day, thus preserving the signature of the first generation 
of stars (e.g. \citealp{Simon2014}). In particular, investigating the metallicity distribution of metal-poor stars, the so-called metallicity distribution 
function (MDF) (see \citealp{Beers2005} for a thorough review), places constraints on several key physical quantities, 
such as the critical metallicity, $Z_{\rm crit}$, and the characteristic 
mass of Pop~III stars. 
The MDF for the metal-poor Milky Way halo stars 
exhibits a peak at $\rm [Fe/H]\approx-2$, and rapidly declines toward lower metallicities with a sharp 
cutoff at $\rm [Fe/H]\approx-4$.
\par
\begin{figure}
\includegraphics[width=80mm]{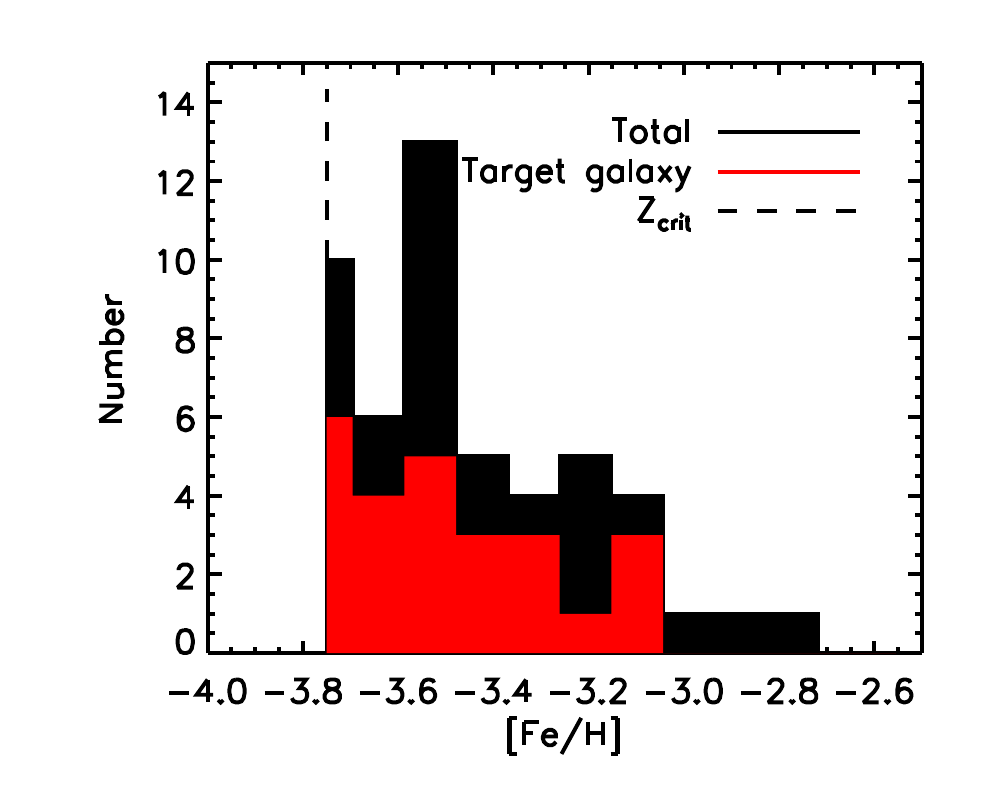}
 \caption{Metallicity distribution function (MDF) of Pop~II stars 
     formed by the end of the simulation. The MDF peaks at $Z_{\rm s}=10^{-3.5}\zsun$ and 
     $Z_{\rm s}=10^{-3.3}\zsun$, respectively, for the stars inside the target halo ({\it red}) 
     and for all Pop~II stars in the high-resolution region ({\it black}). 
The lower cutoff is a consequence of the adopted critical metallicity, which may
be lower in reality.
     \label{Fig:stellar_m}}
\end{figure}

The semi-analytical study by \citet{Salvadori2007}, where a Monte Carlo methodology based on an 
analytic Press-Schechter merger-tree is employed, shows that their model for which $Z_{\rm crit}=10^{-4}\zsun$ and 
$M_{\rm Pop~III, char}=200\msun$ can nicely reproduce the observed Galactic halo MDF, except for the 
existence of hyper metal-poor stars below $\rm [Fe/H]=-5$. Fig.~\ref{Fig:stellar_m} shows the MDF 
of Pop~II stars in our simulation, within the target halo (red) and within the more extended high-resolution region (black), evaluated at the end of our simulation at $z\approx10.5$. We should mention that stellar metallicities, $Z_{\rm s}$, are different from the gas metallicities discussed in previous sections. The stellar metallicity is inherited from the gas metallicity when a star is born out of its birth cloud.
For the conversion between $Z_{\rm s}$ and $\rm [Fe/H]$, we use the relation $\rm [Fe/H]\simeq \log_{\rm 10}(Z_{\rm s}/\zsun)-0.25$, adopting 
the relative iron fraction produced by Pop~III stars (\citealp{Heger2010}). The cut-off at $\rm [Fe/H]=-3.75$ 
arises from the critical metallicity $Z_{\rm crit}=10^{-3.5}\zsun$ adopted here, below which Pop~II stars cannot form. Thus, 
all stellar metallicities exceed $\rm [Fe/H]=-3.75$ by design, and the maximum is 
at $\rm [Fe/H]=-3.75$ and $\rm [Fe/H]=-3.55$ for the target halo and the high-resolution region, respectively. 
\par
The stellar metallicity is determined by a complex interplay between the strength of photoheating by Pop~III stars 
and the ability of haloes to retain their gas. In a halo with a shallow potential or in a halo hosting a massive 
Pop~III star, the stellar metallicity of Pop~II stars tends toward the metallicity 
of the reincorporated gas that was ejected out of the host halo by a previous SN blastwave. On the other hand, 
if photoheating is not strong, as is usually the case for the low mass Pop~III stars ($M_{\rm Pop~III}\lesssim40\msun$), or 
if a host halo is massive, Pop~II stars are likely to be born out of the gas that is highly enriched, because after a SN explosion the 
ejected metals remain in the central region and the central gas density remains high, promptly 
triggering Pop~II star formation.
\par
We should caution that our MDF is not directly comparable to the observed MDF, particularly to the low-metallicity 
tail of the observed MDF ($\rm [Fe/H]<-3.75$). The main reason for the lack of these extremely metal-poor (EMP) stars 
in the simulated MDF is due to the critical metallicity we adopt. Hence, in order to include such stars we might use a 
lower value of the critical metallicity. The other way of explaining EMP stars would be to assume that low mass, 
long-lived metal-free stars ($M_{\rm Pop~III}<0.8\msun$) were formed, and that their surfaces were polluted externally later on
by accretion of gas enriched with metals (e.g. \citealp{Shigeyama2003}; \citealp{Trenti2010}; \citealp{Johnson2010}).
\par
One of the interesting open questions in high-$z$ galaxies is whether or not we can find any fossils of the 
first galaxies in the Local Group, especially in local dwarf galaxies 
(see \citealp{Tolstoy2009} for a review; see also e.g. \citealp{Bovill2009}; \citealp{Simpson2013}; \citealp{Milos2014}). 
One widely discussed scenario for the origin of dwarf galaxies invokes the suppression
of their gas supply due to reionization (e.g. \citealp{Salvadori2009}). Specifically,
most stars are argued to have formed before reionization in haloes with $M_{\rm vir}=10^8-10^9\msun$, giving rise to systems that have then
passively evolved into a subset of the dwarf galaxies. 
Next to reionization, 
photoheating and SN feedback act to evacuate the gas from the haloes. This scenario 
would naturally explain both the old age of the extant stellar populations, and the absence of any ongoing star formation in dwarf galaxies. 
However, the probability of a galaxy formed near the reionization era to remain intact and isolated, not 
undergoing mergers, is small, of order $\approx10\%$ (\citealp{Sasaki1994}). Therefore, 
only a subset of the present-day Local Group dwarfs might be suitable candidates for comparison with 
the first galaxies \citep{Frebel2012,Frebel2014}.
\par
\begin{figure}
\includegraphics[width=80mm]{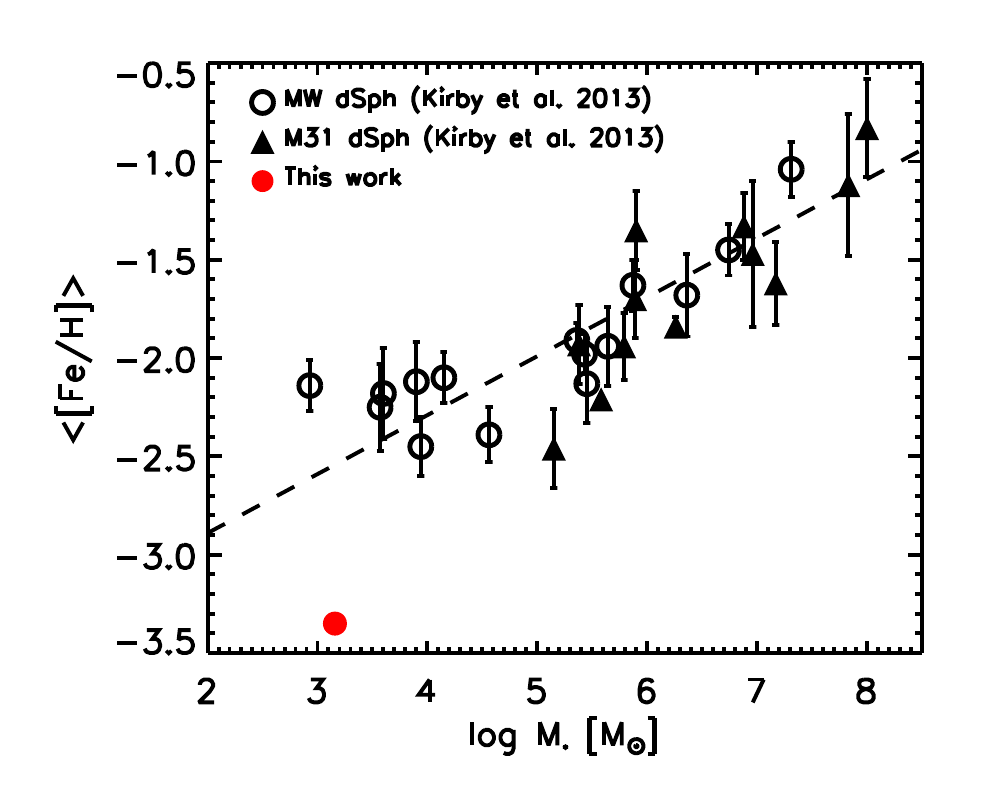}
  \caption{Stellar mass--mean metallicity relation from observations of dwarf galaxies around our Milky Way 
  and M 31. They are strongly correlated, well fitted by the formula
      <$\rm [Fe/H]$>$\approx-1.69+0.30 \log_{10}(M_{\ast}/10^6\msun)$, shown as the dashed line (\citealp{Kirby2013}).
      For comparison, we also plot the estimate for the mean metallicity of our simulated galaxy ({\it red dot}). This value is lower than the empirical relation by $\approx1$ dex, implying that the simulated system cannot give rise to the observed local dwarfs. However, see the main text for further discussion.
        \label{Fig:Mass-metal}}
\end{figure}

Fig.~\ref{Fig:Mass-metal} shows the relation between the mean metallicity of metal-poor stars and their masses 
in dwarf galaxies of the Milky Way and M 31 (\citealp{Kirby2013}). 
Assuming that stellar metallicities remain unchanged until today, the estimated mean metallicity of our simulated galaxy 
is <$\rm [Fe/H]$>$\approx-3.3$, which is lower by $\approx0.5$ dex than what is expected from the empirical relation,
  <$\rm [Fe/H]$>$\approx-1.69+0.30 \log_{10}(M_{\ast}/10^6\msun)$ (see equ.~4 in \citealp{Kirby2013}).
However, our simulation ends before reionization is complete. Indeed, further star formation and metal enrichment will likely occur, leading to Pop~II stars with higher metallicities. The effect of this missing activity may be to raise the average metallicity, possibly up to levels consistent with the empirical Kirby-relation.
Therefore, tracing the
evolution of the galaxy to $z\simeq 5$, and ideally even beyond that, might be necessary to directly compare simulation results with observations.
\par
To infer the star formation history of dwarf galaxies, \citet{Webster2014} have recently performed simulations that follow the evolution of an
isolated system with a dark matter mass of $\sim10^7\msun$ for 600 Myr, accounting for metal enrichment by Type~II and 
Type~Ia SNe. They found that their simulated system has a mean metallicity of $\rm [Fe/H]=-2.1$ and a stellar 
mass of $2\times10^3\msun$ after 600 Myr, suggesting Segue~1 as the closest observed match (see also \citealp{Frebel2014}). 
However, their simulation is performed in a small region (< 400 pc) with a rather idealized setup.
In particular, their high mean metallicity is caused by the assumption of an enrichment
floor that is linearly increasing with the number of SNe. On the other hand, Pop~II SN enrichment in our simulation does not impose any artificial metallicity floor, 
thus enabling lower 
metallicity conditions. 

\begin{figure*}
  \includegraphics[width=130mm]{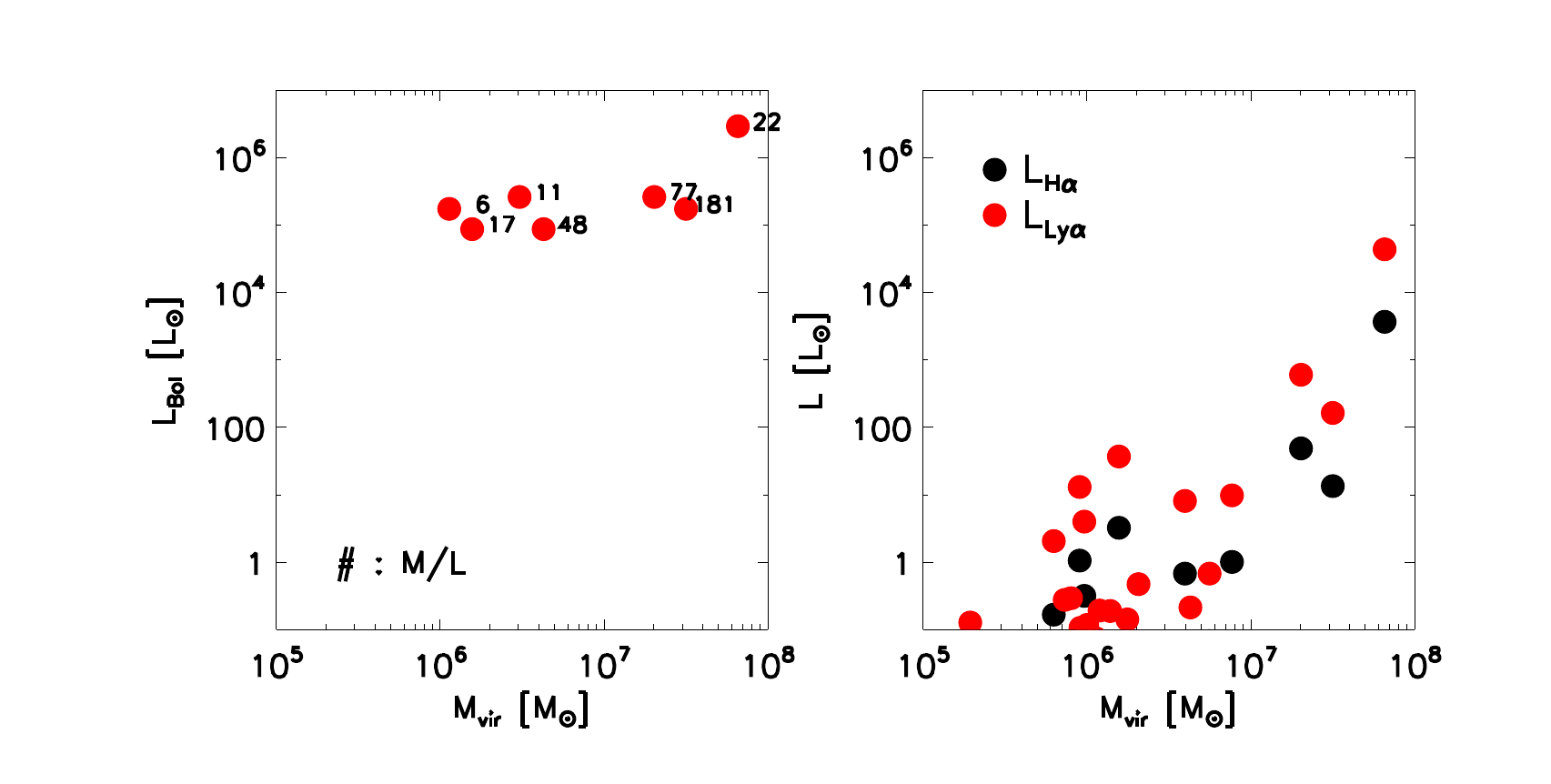}
     \caption{Bolometric luminosity ({\it left}) and recombination line luminosity from $\rm H\alpha$ 
     and $\rm Ly\alpha$ ({\it right}) for haloes hosting Pop~II clusters. The target galaxy has a
     bolometric luminosity of $L_{\rm Bol}=3.5\times10^6\lsun$. Except for this target galaxy, 
     the bolometric luminosities of the other haloes are of order a few times $10^5\lsun$. The numbers next to the 
     red dots in the left panel indicate the mass-to-light ratios, which are quite large, similar to that of the local dwarf 
     galaxies. The recombination line luminosities from $\rm H\alpha$ and $\rm Ly\alpha$ are 
     smaller than $\rm L_{\rm Bol}$ by $1-4$ orders of magnitude. \label{Fig:lum}}
\end{figure*}

\subsection{Luminosity}
One of the imminent challenges for the next generation of telescopes, such as the {\it JWST}, is to detect the first light 
from high-redshift dwarf galaxies in the early Universe. In order to assess the detectability of these 
sources, we compute the expected bolometric and line emission luminosities for the simulated galaxy. 
Due to the short lifetime of Pop~III stars, we only 
consider emission from Pop~II stars. Assuming an instantaneous starburst, the bolometric luminosity 
from a Pop~II cluster is represented by the following equation, 
\begin{equation}
L_{\rm Bol, Pop~II}\propto \int^{m_{\rm max}}_{m_{\rm min}} m^{-\alpha}\phi(m)\hspace{0.1cm}dm,
\end{equation}
where $\phi(m)$ is the mass-luminosity relation.
Depending on the mass of a star, a different power-law is 
used: $L\propto m^{2.3}$ ($m<0.43\msun$), 
$L\propto m^4$ ($0.43\msun<m<2\msun$), $L\propto m^{3.5}$ ($2\msun<m<20\msun$), 
and $L\propto m$ ($m>20\msun$).
Additionally, we compute the nebular emission, i.e. the luminosity in recombination lines, particularly for 
$\rm H\alpha$ and $\rm Ly\alpha$, from the surrounding gas that is ionized by star formation 
(e.g. \citealp{Johnson2009}),
\begin{equation}
L_{\rm H\alpha} = \sum_i j_{\rm H\alpha}\frac{m_i}{\rho_i}\left[ \frac{\rho_i}{\mu_i m_{\rm H}}\right]^2 f_{\rm e} f_{\rm HII},
\end{equation}
\begin{equation}
L_{\rm Ly\alpha} = \sum_i j_{\rm Ly\alpha}\frac{m_i}{\rho_i}\left[ \frac{\rho_i}{\mu_i m_{\rm H}}\right]^2 f_{\rm e} f_{\rm HII},
\end{equation}
where $j$ is the temperature-dependent emission coefficient of the lines (\citealp{Osterbrock2006}), 
and $f_{\rm e}$ and $f_{\rm HII}$ are the fractions of electrons and H~II ions of the SPH particles 
inside the virial volume of a given halo.
\par
Fig.~\ref{Fig:lum} shows that the most massive halo with a mass 
$M_{\rm vir}\sim8\times10^7\msun$, hosting the target galaxy, has a bolometric luminosity of $L_{\rm Bol}=3.5\times10^6\lsun$. 
We note that 20$\%$ of Pop~II clusters (out of 34) are younger than 20 Myr, and these are the main contributors to the 
luminosity. For the other less massive haloes in the mass range of 
$M_{\rm vir}=[1.2\times10^6\msun,3\times10^7\msun]$, the bolometric luminosity varies in the range of 
$L_{\rm Bol}=[8.7\times10^4\lsun, 2.6\times10^5\lsun]$. Our estimates are consistent with those in \citet{Wise2014}, who use a 
more sophisticated stellar population 
model to derive bolometric luminosity.
As stated above, we only consider the luminosity from Pop~II clusters. 
We should note, however, that the simulated target galaxy continues to form Pop~III stars as well, out of the gas with 
metallicity below $Z_{\rm crit}=10^{-3.5}\zsun$. At $z\approx10.5$, 
the galaxy hosts 6 Pop~III stars with a total luminosity of
$1.4\times10^6\lsun$, corresponding to $40\%$ of the Pop~II bolometric luminosity. Thus, adding the contribution 
from Pop~III stars, the combined luminosity may increase roughly by a factor of $\sim$1.4. 
\par
Using our bolometric luminosity estimates, we can
predict the observed flux at a rest-frame 
wavelength of $\lambda_{\rm e}=1500$\AA, characteristic of the soft (non-ionizing) UV continuum.
The observed continuum flux for an unresolved object is given by (e.g. \citealp{Oh1999}; \citealp{Bromm2001a})
\begin{eqnarray}
f_{\nu}(\lambda_{\rm o}) &=& \frac{L_{\nu}(\lambda_{\rm e})}{4\pi d_{\rm L}^2(z)} (1+z) \\
&\sim& 1\unit{nJy} \left( \frac{L_{\nu}(\lambda_{\rm e})}{10^{27} \unit{erg} \unit{s^{-1} Hz^{-1}}} \right) \left( \frac{1+z}{11} \right)^{-1},
\end{eqnarray}
where $d_{\rm L}$ is the luminosity distance to a source at redshift $z$, which is $\sim100$ Gpc at $z=10$. The predicted flux for 
the galaxy ($M_{\rm vir}=8\times10^7\msun$) in our simulation is $\approx10^{-3}\unit{nJy}$ for $L_{\rm Bol}=3.5\times10^6\lsun$. 
This is smaller by two orders of magnitude than the detection limit of the Near Infrared Camera (NIRCam) 
onboard the {\it JWST} for extremely deep imaging
(e.g. \citealp{Gardner2006}). Therefore, more
promising targets for detection with the {\it JWST} will be more massive, luminous, systems with
$M_{\rm vir}=10^9-10^{10}\msun$ (e.g. \citealp{Pawlik2011}).
The other possible way of detecting those small galaxies is through gravitational lensing with a boost in brightness 
by a factor of $\mu\sim100$ \citep[e.g.][]{Zackrisson2012, Zackrisson2014}. However, magnifications of $\mu>100$ 
are extrmely rare, mainly due to the low probability of having sight lines for the gravitational lensing, 
$P(>\mu)\propto\mu^{-2}$ (e.g. \citealp{Pei1993}), requiring very large survey areas to find the lensed galaxies.
\par
On the other hand, the nebular emission in recombination lines are 
$L_{\rm H\alpha}=4\times10^3\lsun$ and $L_{\rm Ly\alpha}=4.8\times10^4\lsun$, respectively, 
both of which are even weaker than the bolometric luminosity by $2-3$ orders of magnitude. This is mainly due to the low gas densities as a result of the
photoheating from the newly forming stars, resulting in low recombination rates. 
In our estimate, we have assumed escape fractions for $\rm H\alpha$ and $\rm Ly\alpha$ photons of unity, rendering 
the calculated line luminosities upper limits. Although the recombination luminosity in $\rm Ly\alpha$ is higher 
than that in $\rm H\alpha$, in reality $\rm Ly\alpha$ photons are more likely to scatter in the neutral IGM 
(e.g. \citealp{Dijkstra2007}; \citealp{Smith2014}), and thus the 
observed flux in $\rm Ly\alpha$ may be substantially lower than implied by the intrinsic luminosity.  
Because of the hard spectra of massive Pop~III stars, strong nebular emission, particularly in the He~II $\lambda1640$ line, has 
been suggested as a possible observable signature of metal-free Pop~III galaxies 
(e.g. \citealp{Schaerer2003}; \citealp{Dawson2004}; \citealp{Nagao2008}; \citealp{Johnson2009}; \citealp{Cai2014}). 
However, as implied by our simulation, the gas inside the first galaxies is likely already 
metal-enriched, and thus their stellar content is expected to be dominated by Pop~II, instead of Pop~III, stars, making 
it more difficult to use the recombination lines as a probe to detect such metal-enriched small systems.
\par
Finally, a complementary method is to look for counterparts of the first galaxies in the Local Group. Assuming, probably unrealistically so, that star formation 
in our simulated galaxy is completely quenched after $z\approx10.5$, and that it has passively evolved until the present day, 
the total stellar mass is $1.4\times10^3\msun$ at $z=0$ for the IMF we use. This is comparable to the stellar mass $M_{\ast}=900\msun$ of Segue~2, 
the least luminous dwarf galaxy known in the local Universe (\citealp{Kirby2013a}). 
More high-resolution observations of the faintest galaxies are needed to further elucidate the connection between the 
fossils of the first galaxies and local dwarf galaxies.

\section{Summary and discussion}
We have carried out a radiation hydrodynamical simulation of a galaxy reaching a mass of $M_{\rm vir}=8\times10^7\msun$ 
at $z\approx10$. This simulation tracked ionizing photons from individual Pop~III stars 
and from Pop~II clusters. High spatial resolution allowed us to explicitly account for the mechanical and chemical feedback 
from CCSNe and PISNe. Additionally, we followed the transport
of X-ray photons from accreting black holes (BHs) and high mass X-ray binaries (HMXBs). Motivated by recent works on primordial 
star formation, which suggest Pop~III stars with a few tens of $\msun$, we adopted a characteristic mass, 
$M_{\rm char}=20\msun$ for the Pop~III IMF. This is substantially smaller than $M_{\rm char}=100\msun$, which has 
been used in most previous studies. We also investigated the metal enrichment history in the early Universe, including 
the transition in star formation mode from metal-free Pop~III stars 
to low-mass, metal-enriched Pop~II stars at the critical metallicity $Z_{\rm crit}=10^{-3.5}\zsun$. Our main results are as follows.
\\
\par
(i) Adopting a moderate characteristic mass $M_{\rm char}=20\msun$, instead of $M_{\rm char}=100\msun$, for the Pop~III IMF, results in a 
substantially different star formation history during the assembly process of the target halo that hosts the first galaxy. 
This is caused by the relatively weak stellar feedback, particularly photoheating, from the less massive Pop~III stars.
\\
\par
(ii) The gas inside the galaxy is enriched to $Z_{\rm avg}\sim10^{-3}\zsun$ by a total of 40 Pop~III and 
248 Pop~II core-collapse supernovae (CCSNs). Despite the striking difference in the star formation history, 
the degree of metal enrichment by CCSNe is comparable to that by a single 
$200\msun$ Pop~III pair-instability supernova (PISN).
\\
\par
(iii) The metals produced by two PISNe outside the main galaxy are mainly responsible for the metal enrichment 
in the diffuse IGM, setting a metallicity floor of $Z=10^{-4}\zsun$. Therefore, signatures of PISNe in the early Universe may 
be found by studying the state of the IGM, e.g. employing bright Pop~III GRB afterglows.
\\
\par
(iv) There is no gas with metallicity $Z\lesssim10^{-5}\zsun$ inside the main galaxy. Given the critical metallicity 
$Z_{\rm crit}=10^{-3.5}\msun$ adopted here, Pop~III stars continue to form out of the gas in the metallicity range 
$10^{-5}\zsun \lesssim Z \lesssim 10^{-3.5}\zsun$. Thus, both Pop~III and Pop~II stars coexist.
This conclusion, however, is uncertain, as the critical metallicity may well be lower,
which would not allow any extended Pop~III star formation.
\\
\par
(v) Due to the rarity of HMXBs and their short lifetimes, in combination with the low BH accretion 
rates $\dot{M}_{\rm BH}=10^{-9}-10^{-12}\msun \unit{yr^{-1}}$, the impact of X-rays on the gas 
inside haloes that host X-ray sources and on the IGM is negligible in this simulation.
\\
\par
(vi) The bolometric luminosity of the simulated galaxy is $L_{\rm Bol}\approx3.5\times10^6\lsun$ and the 
total stellar mass is $6\times10^3\msun$ at $z\approx10.5$. Therefore, the UV flux of the galaxy, redshifted into the near-IR is 
$1.4\times10^{-3} \unit{nJy}$, significantly less than the detection limit for NIRCam onboard the {\it JWST}.
\\
\par
(vii) The estimated metal distribution function (MDF) of Pop~II stars in the galaxy peaks at $\rm [Fe/H]=-3.75$ and the stellar metallicities 
are in the range $\rm [Fe/H]=[-3.75, -2.7]$ with a mean metallicity of $\rm [Fe/H]=-3.3$. 
The stellar mass of $M_{\ast}=1.4\times10^3\msun$ of the simulated galaxy, passively evolved to the present day, is comparable 
to that of Segue~2, the least luminous dwarf known in the local Universe.
\\
\par
It has often been pointed out that including a physically motivated description of stellar feedback is 
crucial for modeling galaxy formation, especially for small galaxies, which are susceptible to stellar 
feedback due to their shallow potential wells (e.g. \citealp{Wise2012}; \citealp{Kim2013}; \citealp{Hopkins2013}). 
We confirm that stellar feedback plays a crucial role in shaping the properties of the first galaxies: by 
reducing the baryon fraction ($f_{\rm bar}\approx0.05$), corresponding to $15\%$ of the cosmic mean,
and by enriching the gas to an average of <$Z$> $\approx10^{-3}\zsun$, making Pop~II star formation the dominant mode inside the first galaxies.
However, galaxies may continue to form Pop~III stars because not all gas is 
enriched above the critical metallicity.

In our simulation, we needed to fix a large number of numerical input 
parameters, such as the random sampling of the Pop~III stellar masses
from the underlying IMF. Given that computational resources limit us
to only one such realization from a very large ensemble of likely
parameter choices, the question arises what aspects of our simulation
are contingent on the specific choice, and which results are likely robust.
An example of a robust feature is that the first galaxies were already
metal-enriched, such that their stellar content is expected to be Pop~II; 
the precise level of this `bedrock metallicity', on the other hand, is
uncertain, but likely of order
$Z_{\rm avg}>10^{-4}\zsun$ at $z\approx10$. An example of a contingent,
numerical-parameter dependent feature is the impact of X-ray feedback from
early HMXB sources.
We mention that X-ray feedback is negligible in our simulation, but the population of HMXBs, which are likely the strongest X-ray sources, 
and their luminosity and duty-cycle are unknown at high redshifts. We refer the reader to \citet{Jeon2012} for discussion of a case where X-ray feedback is more significant.
\par

We show that the stellar mass fraction in halos is $f_{\ast}=M_{\ast}/M_{\rm vir}\sim 10^{-4}$, which is smaller by an oder of magnitude than the 
values found in other studies, for example \citet{Behroozi2013} and \citet{OShea2015}, using abundance-matching techniques and grid-based simulations, respectively. However, the scatter from their work is large, such that $f_{\ast}$ ranges from $10^{-5}$ to $10^{-1}$ at a given mass of $M_{\rm vir}=10^8\msun$. On the other hand, our estimated for $f_{\ast}$ is in a good agreement with the results from \citet{GK2014}, in which they use an updated abundance-matching technique and reproduce the stellar mass functions of the Milky Way and Andromeda satellites more accurately. The fact that near the end of our simulation the baryon fraction of the target halo is significantly reduced to 15$\%$ of the cosmic mean, leading to a reduction in star formation, robustly implies a highly feedback-regulated regime. Such inefficiency of star formation could be attributed to how we implement stellar feedback, especially the efficiency of SN feedback. However, even if the stellar mass were increased by a factor of 10 by adopting a different parameter implementation, this system would still be observationally out of reach for the {\it JWST}, at least in the absence of gravitational lensing.

\par
Whether fine-structure line cooling or dust-continuum cooling was mainly responsible 
for Pop~II star formation is still debated (e.g. \citealp{Omukai2000}; \citealp{Bromm2001b};
\citealp{Schneider2002}; \citealp{BrommLoeb2003};
\citealp{Omukai2005}; \citealp{Schneider2010}; \citealp{Schneider2012};
 \citealp{Chiaki2014}; \citealp{Ji2014}). If we had adopted a critical metallicity $Z_{\rm crit,dust}=10^{-5.5}\zsun$, 
 set by dust continuum cooling, the assembly history of the target galaxy would have proceeded quite differently. Adopting a lower critical metallicity than $Z_{\rm crit}=10^{-3.5}\zsun$ would boost the relative importance of Pop~II star formation, and long-lived Pop~II stars could hence contribute to the total stellar mass. The estimated MDF of Pop~II stars would also be changed depending on the critical metallicity. The nature and observational properties of the first galaxies thus sensitively depend not only on the mass scale of the first stars, as has previously been pointed out \citep{Bromm2011}, but also on additional astrophysical parameters, such as the critical metallicity.
 \par 
Considerable uncertainties remain in the treatment of metal mixing. We used 
a subgrid model (\citealp{Greif2009}) in which we assume 
that the efficiency of metal mixing below the resolved scales is determined by physical quantities at 
resolved scales. \citet{Shen2010} propose a 
new implementation to more accurately capture the turbulent mixing in SPH. Further theoretical and observational 
studies are required to better understand the metal mixing process. As discussed, exploring a large range of parameters is needed to derive representative characteristics of high-$z$ dwarf galaxies. However, performing such a comprehensive suit of high-resolution, radiation-hydrodynamical simulations is computationally challenging.

As mentioned above, the direct detection of the simulated galaxy, with $M_{\rm vir}\lesssim10^8\msun$, will be out of reach even for the {\it JWST}, unless
extreme, and probably unattainable, gravitational lensing magnifications are involved. Therefore, a more promising strategy 
to infer their nature may be to use ``stellar" or ``galactic" archeology by searching for metal-poor stars in the Milky 
Way or in nearby dwarf galaxies. To fully unleash its potential, both simulations
and observational surveys need to be pushed further, a process that clearly has been
set in motion now.



\section*{acknowledgements}
We are grateful to Volker Springel, Joop Schaye, and Claudio Dalla
Vecchia for letting us use their versions of \textsc{gadget} and their data
visualization and analysis tools. We thank Jun-Hwan Choi for discussions 
on the simulation results. V.~B.\ and M.~M.\ acknowledge support
from NSF grants AST-1009928 and AST-1413501.
A.~H.~P.\ receives funding from the European Union's Seventh Framework
Programme (FP7/2007-2013) under grant agreement number
301096-proFeSsoR. We acknowledge access to compute resource Supermuc based in Germany at LRZ Garching provided by the Gauss Centre for Supercomputing/Leibniz Supercomputing Centre under grant:pr83le and by PRACE (proposal number 2013091919). The authors acknowledge the Texas Advanced 
Computing Center at The University of Texas at Austin for providing 
HPC resources under XSEDE allocation TG-AST120024
\footnotesize{

\bibliography{myrefs2}{}
\bibliographystyle{mn2e}

}
 
\end{document}